\def\spose#1{\hbox to 0pt{#1\hss}}
\def\simlt{\mathrel{\spose{\lower 3pt\hbox{$\mathchar"218$}}
     \raise 2.0pt\hbox{$\mathchar"13C$}}}
\def\simgt{\mathrel{\spose{\lower 3pt\hbox{$\mathchar"218$}}
     \raise 2.0pt\hbox{$\mathchar"13E$}}}
\def\simpropto{\mathrel{\spose{\lower 3pt\hbox{$\mathchar"218$}}
     \raise 2.0pt\hbox{$\propto$}}}

\newcommand{\appropto}{\mathrel{\vcenter{
  \offinterlineskip\halign{\hfil$##$\cr
    \propto\cr\noalign{\kern2pt}\sim\cr\noalign{\kern-2pt}}}}}

\documentclass[twocolumn,aps,prd,nofootinbib,showpacs]{revtex4-1}
\usepackage{amsmath,graphicx,bm,color}
\begin{document}
\title{The Epoch of Reionization Window: II. Statistical Methods for Foreground Wedge Reduction}

\author{Adrian Liu}
\email{acliu@berkeley.edu}
\affiliation{Department of Astronomy, UC Berkeley, Berkeley, CA 94720, USA}
\affiliation{Berkeley Center for Cosmological Physics, UC Berkeley, Berkeley, CA 94720, USA}
\author{Aaron R. Parsons}
\affiliation{Department of Astronomy, UC Berkeley, Berkeley, CA 94720, USA}
\affiliation{Radio Astronomy Laboratory, UC Berkeley, Berkeley, CA 94720, USA}
\author{Cathryn M. Trott}
\affiliation{International Centre for Radio Astronomy Research, Curtin University, Bentley, WA, Australia}
\affiliation{ARC Centre of Excellence for All-Sky Astrophysics (CAASTRO), Curtin University, Bentley WA, Australia}
\date{\today}

\newcommand{\apjs}{Astrophys. J. Suppl. Ser.}
\newcommand{\aj}{Astron. J.}
\newcommand{\mnras}{Mon. Not. R. Astron. Soc.}
\newcommand{\apjl}{Astrophys. J. Lett.}
\newcommand{\aap}{Astron. Astrophys.}
\newcommand{\pasa}{PASA}
\newcommand{\physrep}{Phys. Rep.}
\newcommand{\araa}{Annu. Rev. Astron. Astrophys.}


\begin{abstract}
For there to be a successful measurement of the $21\,\textrm{cm}$ Epoch of Reionization (EoR) power spectrum, it is crucial that strong foreground contaminants be robustly suppressed.  These foregrounds come from a variety of sources (such as Galactic synchrotron emission and extragalactic point sources), but almost all share the property of being spectrally smooth, and when viewed through the chromatic response of an interferometer, occupy a signature ``wedge" region in cylindrical $k_\perp k_\parallel$ Fourier space.  The complement of the foreground wedge is termed the ``EoR window", and is expected to be mostly foreground-free, allowing clean measurements of the power spectrum.  This paper is a sequel to a previous paper that established a rigorous mathematical framework for describing the foreground wedge and the EoR window.  Here, we use our framework to explore statistical methods by which the EoR window can be enlarged, thereby increasing the sensitivity of a power spectrum measurement.  We adapt the Feldman-Kaiser-Peacock approximation (commonly used in galaxy surveys) for $21\,\textrm{cm}$ cosmology, and also compare the optimal quadratic estimator to simpler estimators that ignore covariances between different Fourier modes.  The optimal quadratic estimator is found to suppress foregrounds by an extra factor of $\sim 10^5$ in power at the peripheries of the EoR window, boosting the detection of the cosmological signal from $12\sigma$ to $50\sigma$ at the midpoint of reionization in our fiducial models.  If numerical issues can be finessed, decorrelation techniques allow the EoR window to be further enlarged, enabling measurements to be made deep within the foreground wedge.  These techniques do not assume that foreground are Gaussian-distributed, and we additionally prove that a final round of foreground subtraction can be performed after decorrelation in a way that is guaranteed to have no cosmological signal loss.
\end{abstract}

\maketitle
\section{Introduction}

By mapping the intensity of the redshifted hyperfine transition of hydrogen in three dimensions, $21\,\textrm{cm}$ cosmology has the potential to survey a larger volume of our observable Universe than any other cosmological probe to date \cite{Furlanetto2006,Morales2010,Pritchard2012,AviBook}.  Efforts aimed at lower redshifts ($0 <  z \lesssim 4$, depending on the experiment) aim to use neutral hydrogen as a tracer for large scale structure and are expected to be incisive probes of dark energy and its time evolution \cite{Pober2013a,Ansari2012,Battye2012,Shaw2014a,SaiyadAli2013,Chen2012}.  Interferometer arrays such as the Donald C. Backer Precision Array for Probing the Epoch of Reionization (PAPER \cite{Parsons2010,Parsons2013}), the Murchison Widefield Array (MWA \cite{Tingay2013,Bowman2013}), the Low Frequency Array (LOFAR \cite{Yatawatta2013}), and the Giant Metrewave Radio Telescope Epoch of Reionization experiment (GMRT-EoR \cite{Paciga2013}) are currently targeting higher redshifts ($6 \lesssim z \lesssim 13$, again depending on the experiment).  These experiments open up a previously unexplored period in our Universe's history---the Epoch of Reionization (EoR)---during which the first luminous objects ionized the intergalactic medium (IGM).  Future instruments such as the Hydrogen Epoch of Reionization Array (HERA \cite{Pober2014}) and the Square Kilometer Array (SKA \cite{Mellema2013}) will further push the redshift and sensitivity frontiers, probing an even broader range of redshifts with greater sensitivity.  This will enable not only detailed studies of the properties of the first objects and their effect on the IGM, but also of more exotic physics such as dark matter annihilation \cite{Valdes2013}, and may eventually even lead to constraints on fundamental parameters such as the neutrino mass \cite{McQuinn2006,Mao2008} and non-Gaussianity \cite{Mao2013}.

For $21\,\textrm{cm}$ cosmology to become a reality, however, it will be necessary to deal with foreground systematics.  The cosmological signal that experiments seek to detect is faint (favored to be roughly a few mK in brightness temperature with most theoretical models \cite{Lidz2007,Barkana2009,Zahn2011}), whereas sources of foreground radio emission such as Galactic synchrotron radiation are bright (known empirically to be on the order of a few hundred Kelvin at frequencies relevant to EoR experiments \cite{deOliveiraCosta2008}).  A systematic way to subtract or evade these foregrounds is therefore crucial.

Proposals for foreground mitigation can be roughly split into two categories: foreground subtraction and foreground avoidance.  Foreground subtraction schemes typically propose to model the foregrounds before subtracting them off directly from the data.  Different proposals require modeling the foregrounds to different levels of detail.  Some rely only on the fact that the foregrounds are expected to be spectrally smooth compared to the cosmological signal \cite{Wang2006,Gleser2008,Liu2009a,Bowman2009,Harker2009,Liu2009b,Petrovic2011,Cho2012,Liu2012,Parsons2012b}.  Others also take advantage of the angular dependence of foregrounds \cite{Paciga2011,Liu2011,Dillon2013,Masui2013,Dillon2014}, or leverage their inherent non-Gaussianity \cite{Chapman2012,Chapman2013}.  If modeling inaccuracies can be minimized and accidental cosmological signal losses (if any) are correctly accounted for, foreground subtraction techniques in principle allow extremely high signal-to-noise measurements.

As an alternative to foreground subtraction, one may instead opt for a more conservative approach of foreground avoidance.  Foreground avoidance strategies have been developed primarily for measurements of the $21\,\textrm{cm}$ power spectrum, which will be the focus of this paper.  In a power spectrum measurement, one essentially takes a three-dimensional survey volume and performs spatial Fourier transforms over all three spatial axes to give a set of Fourier amplitudes labeled by three wavenumbers: $k_x$ and $k_y$ for the angular directions, and $k_\parallel$ for the line-of-sight direction.  The data are then squared and appropriately binned.  Now, recall that because $21\,\textrm{cm}$ surveys map our Universe using a spectral line, there is a direct correspondence between the observation frequency and the line-of-sight distance.  Foreground emission (being spectrally smooth \cite{Oh2003}) is therefore expected to contaminate only the lowest $k_\parallel$ modes, unlike the cosmological signal, which is typically spread throughout Fourier space.  It should therefore be possible to pursue a strategy of foreground avoidance, where power spectrum measurements are only made at higher $k_\parallel$.


\begin{figure}[!ht] 
	\centering 
	\includegraphics[width=.49\textwidth]{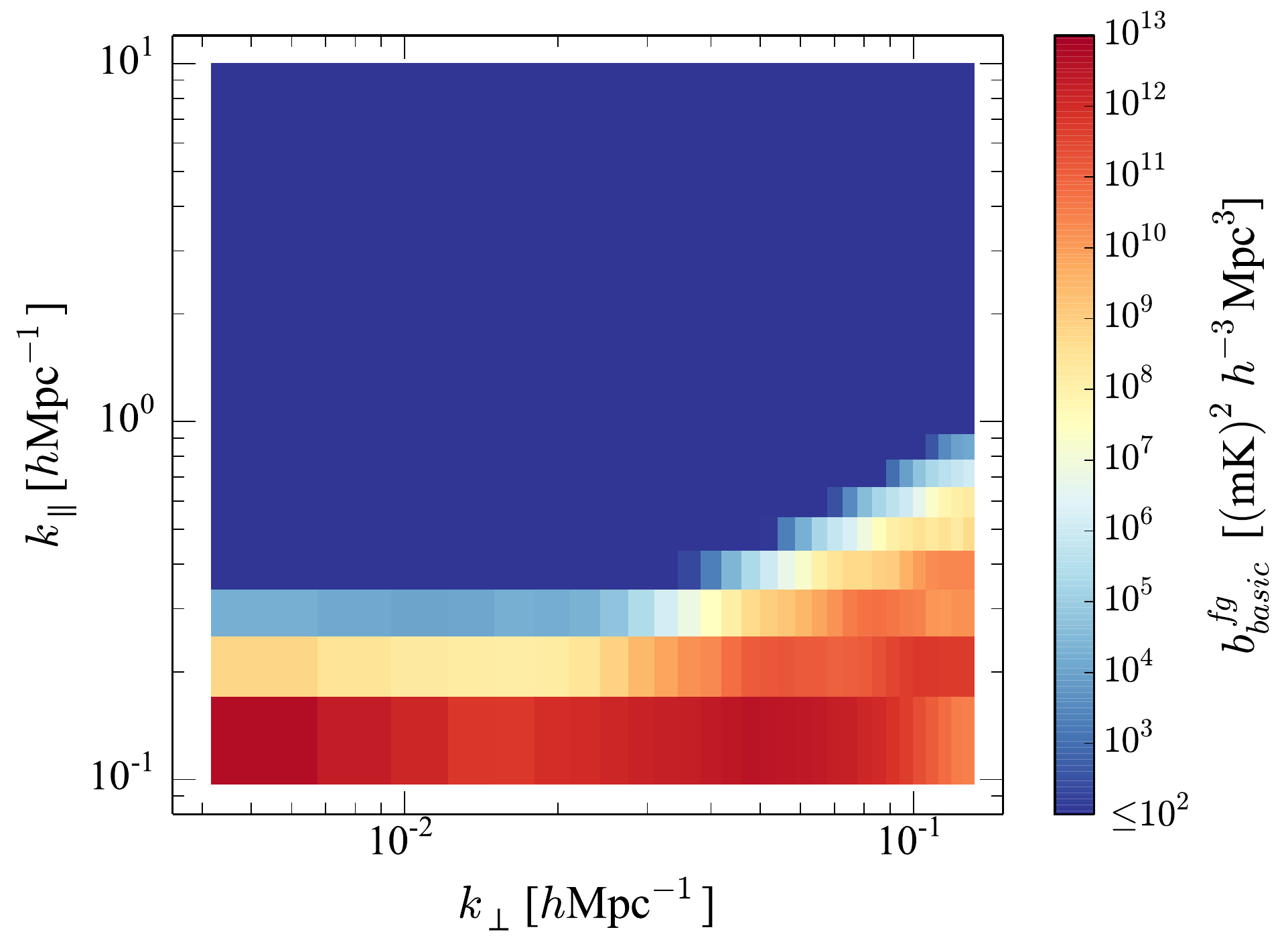}
	\caption{Residual bias (which is expected to contain only foregrounds) for a basic power spectrum estimator, reproduced from Paper I.  The foregrounds strongly contaminate a wedge region, but fall sharply beyond the wedge.}
	\label{fig:basicEstBias}
\end{figure} 

Unfortunately, recent work has shown that it may be overly simplistic to assume that foregrounds occupy only the lowest $k_\parallel$ modes.  In Figure \ref{fig:basicEstBias} we show a numerical computation from a preceding companion paper (Ref. \cite{Liu2014a}, henceforth ``Paper I"), where we plot the expected power spectrum contamination from foregrounds as a function of cylindrical Fourier wavenumbers $k_\parallel$ and $k_\perp \equiv \sqrt{k_x^2 + k_y^2}$.  At low $k_\perp$, we see that our previous expectations are met, with the bright foregrounds contaminating only the first few $k_\parallel$ bins.  At high $k_\perp$, however, the foregrounds extend to higher $k_\parallel$, forming a characteristic ``wedge".  Through detailed simulations, theoretical work, and observations, it is now understood that the wedge arises from the fundamentally chromatic nature of interferometry \cite{Datta2010,Vedantham2012,Parsons2012b,Morales2012,Trott2012,Thyagarajan2013,Hazelton2013,Pober2013b,Dillon2014}.  Since interferometric fringe patterns depend on the observation frequency, the spatial wavenumber $k_\perp$ is coupled to the frequency/line-of-sight wavenumber $k_\parallel$.  This coupling is more pronounced at higher $k_\perp$, which tend to be probed by the baselines of an interferometer array that are longer, and are thus known to be more chromatic.  The result is a leakage of foregrounds from low to high $k_\parallel$ that is more pronounced at high $k_\perp$, i.e., the wedge.

Previous work has shown that while the existence of the wedge means that more of the Fourier plane is contaminated with foregrounds than one might have naively expected, this contamination is limited, in the sense that there are theoretical and observational reasons for a sharp drop-off in foreground power beyond the wedge.  Foreground avoidance is therefore still a viable foreground mitigation strategy, as long as one is careful to only include power spectrum measurements outside the wedge, in a region colloquially known as the ``EoR window".  In Paper I, we extended and unified previously disparate aspects of the existing literature in a mathematical framework of the wedge and the EoR window.  Importantly, our framework was fully covariant, allowing errors and error correlations to be properly captured, whether they arose from instrumental effects or foreground contamination.

In this paper, we make use of our mathematical framework to examine ways in which the EoR window may be enlarged using statistical methods.  In other words, we consider ways in which the foreground wedge can be reduced without directly modeling and subtracting foreground models from the data, thus preserving the conservative spirit of foreground avoidance.  If successful, an enlargement of the EoR window can result in higher significance measurements of the EoR power spectrum, since thermal noise is typically independent of $k_\parallel$, while the cosmological signal increases in strength towards low $k_\parallel$.  The ability to push into the wedge can be the difference between a non-detection and a significant detection with current-generation experiments, and much-improved astrophysical constraints with next-generation experiments \cite{Pober2014}.  We will find that minimum-variance power spectrum estimators can reduce foregrounds at the edge of the wedge by a factor of up to $10^5$ in power (i.e., in temperature-squared units).

Pushing beyond a simple application of the minimum-variance power spectrum estimator, the error statistics computed using the framework presented in Paper I can be used to push deeper into the wedge.  In Paper I, we described the wedge as a scattering of foregrounds from low $k_\parallel$ regions to higher $k_\parallel$.  The form of this scattering is computable in a foreground-independent manner using the formalism of Paper I.  Since this scattering is a well-defined linear operation that acts on the power spectrum, it can in principle be undone using decorrelation techniques.  Decorrelation allows foregrounds to be more readily identified and removed in a final post-processing step following power spectrum estimation.  We prove that such a strategy does not require assumptions of foreground Gaussianity, is immune to possible modeling inaccuracies of the subtracted foregrounds, and does not suffer from any formal signal loss.  The last two features, in particular, stand in contrast to methods that directly subtract foregrounds from the input data \cite{Wang2006,Liu2009a,Bowman2009,Harker2009,Liu2009b,Liu2012}.  With decorrelation and a subsequent foreground removal (a combination that we dub ``foreground isolation"), one may potentially allow work deep within the wedge, although numerical issues must be dealt with.

The rest of this paper is organized as follows.  In Section \ref{sec:PaperI} we establish notation and summarize the results of Paper I.  These results form a basic picture of the wedge and the EoR window that serve as a reference; our goal is to improve upon this basic picture.  Section \ref{sec:Direct} comments briefly on the interplay between foreground subtraction and foreground avoidance.  In Section \ref{sec:Enlarging}, we first examine the possibility of enlarging the EoR window using the computationally cheap class of separable power spectrum estimators before discussing the more computationally expensive non-separable estimators, which include the optimal minimum-variance estimator.  The computationally expensive estimators are explored numerically, using exactly the same setup as we used in Paper I, from the foreground model and the fiducial instrument down to algorithmic details such as binning.  Finally, we examine the foreground isolation approach in Section \ref{sec:Decorr}, before ending with some concluding remarks in Section \ref{sec:Conclusions}.

\section{Summary of Paper I}
\label{sec:PaperI}

In this section, we briefly summarize the results of Paper I \cite{Liu2014a} to set the stage for the estimator optimizations performed in this paper.  In Paper I, we showed how to incorporate the foreground wedge into a discretized quadratic estimator formalism.  The power spectrum is approximated as being comprised of piecewise constant bandpowers in a discretized grid of cells on the $k_\perp k_\parallel$ plane.\footnote{While the cosmological power spectrum is expected to respect statistical isotropy, thus allowing a spherically-symmetric binning of Fourier space along contours of constant $k\equiv (k_\perp^2 + k_\parallel^2)^{1/2}$, the cylindrical power spectrum is the most useful for diagnosing systematics such as foregrounds.  We therefore exclusively deal with the cylindrical power spectrum in this paper, since our focus is on foreground systematics, and one can always include an extra binning step at the end of a power spectrum estimation pipeline to bin the cylindrical power spectrum into a spherical one if desired.}  One then forms an estimate $\widehat{p}_\alpha$ of the bandpower $p_\alpha$ in the $\alpha$th cell on the $k_\perp k_\parallel$ grid by computing
\begin{equation}
\label{eq:GenericEst}
\widehat{p}_\alpha = \mathbf{x}^\dagger \mathbf{E}^\alpha \mathbf{x},
\end{equation}
with $\mathbf{x}$ denoting the input data (e.g. a serialized list of visibilities over all baselines and all frequencies) and $\mathbf{E}^\alpha$ denoting an estimator matrix for the $\alpha$th band.  The form of the estimator matrix is a choice that is made by the data analyst.  Different choices yield estimated bandpowers that are related to the true power spectrum in different ways, with the exact relation specified by window functions.  The window function matrix is defined as
\begin{equation}
\label{eq:WindDef}
\widehat{p}_\alpha \equiv \sum_\beta W_{\alpha \beta} p_\beta,
\end{equation}
so that the $\alpha$th row tells us which linear combination of bands (indexed by $\beta$) contribute to the estimate of the $\alpha$th bandpower.  To allow us to interpret each estimated bandpower as a weighted average of the truth, the normalization of $\mathbf{E}^\alpha$ is chosen such that each row of the window matrix sums to unity.  The explicit form of $W_{\alpha \beta}$ is given by
\begin{equation}
\label{eq:WindExplicitForm}
W_{\alpha \beta} = \textrm{tr} \left[ \mathbf{E}_\alpha \mathbf{C}_{,\beta} \right],
\end{equation}
where $\mathbf{C} \equiv \langle \mathbf{x} \mathbf{x}^\dagger \rangle$ is the total covariance matrix of the data (including foregrounds, instrumental noise, and cosmological signal), and $\mathbf{C}_{,\beta} \equiv \partial \mathbf{C} / \partial p_\beta$ is its derivative with respect to the $\beta$th bandpower.  Throughout this paper, angular brackets denote ensemble averages.  Since we are assuming that the bandpowers are constant within each Fourier cell, the bandpowers are related to the covariance matrix via
\begin{equation}
\label{eq:CovarDecomp}
\mathbf{C} = \mathbf{N} + \sum_\alpha p_\alpha \mathbf{C}_{,\alpha},
\end{equation}
where $\mathbf{N}$ is the instrumental noise covariance

The choice of estimator affects not only the window functions, but also the error bars and the error correlations on the final estimates.  These error statistics are quantified by the bandpower covariance matrix:
\begin{equation}
\Sigma_{\alpha \beta} \equiv \langle \widehat{p}_\alpha \widehat{p}_\beta \rangle - \langle \widehat{p}_\alpha \rangle \langle \widehat{p}_\beta \rangle = 2 \textrm{tr} \left[ \mathbf{E}_\alpha \mathbf{C} \mathbf{E}_\beta \mathbf{C} \right].
\end{equation}
A final quantity in our suite of statistics is the bias in our estimator.  Typically, the bandpower estimates given by Eq. \eqref{eq:GenericEst} will be biased, since the \emph{power} from residual errors will always be positive and therefore not cancel out with averaging, even if the residuals themselves are symmetrically distributed about zero.  The bias $b^\alpha$ for the $\alpha$th bandpower estimate is given by
\begin{equation}
\label{eq:bias}
b_\alpha = \textrm{tr}  [\mathbf{E}^\alpha \mathbf{C}_\textrm{junk}],
\end{equation}
where $\mathbf{C}_\textrm{junk}$ is the contaminant contribution to the data covariance matrix.  In certain applications where this can be accurately modeled (such as in galaxy surveys, where $\mathbf{C}_\textrm{junk}$ is dominated by galaxy shot noise), this bias can be subtracted from the bandpower estimate.  In $21\,\textrm{cm}$ cosmology, the main contribution to $\mathbf{C}_\textrm{junk}$ is the contribution from foregrounds (assuming that noise bias can be eliminated via cross-correlations), which are poorly understood at low frequencies.  We therefore assume that no foreground bias subtraction is attempted, and consider the bias to be a residual systematic.  Put another way, Eq. \eqref{eq:WindDef} does not quite describe the relation between our bandpower estimates and the true bandpowers without the bias subtraction, and needs to be supplemented by Eq. \eqref{eq:bias} as an additive term on the right hand side.

\begin{figure}[!ht] 
	\centering 
	\includegraphics[width=0.49\textwidth]{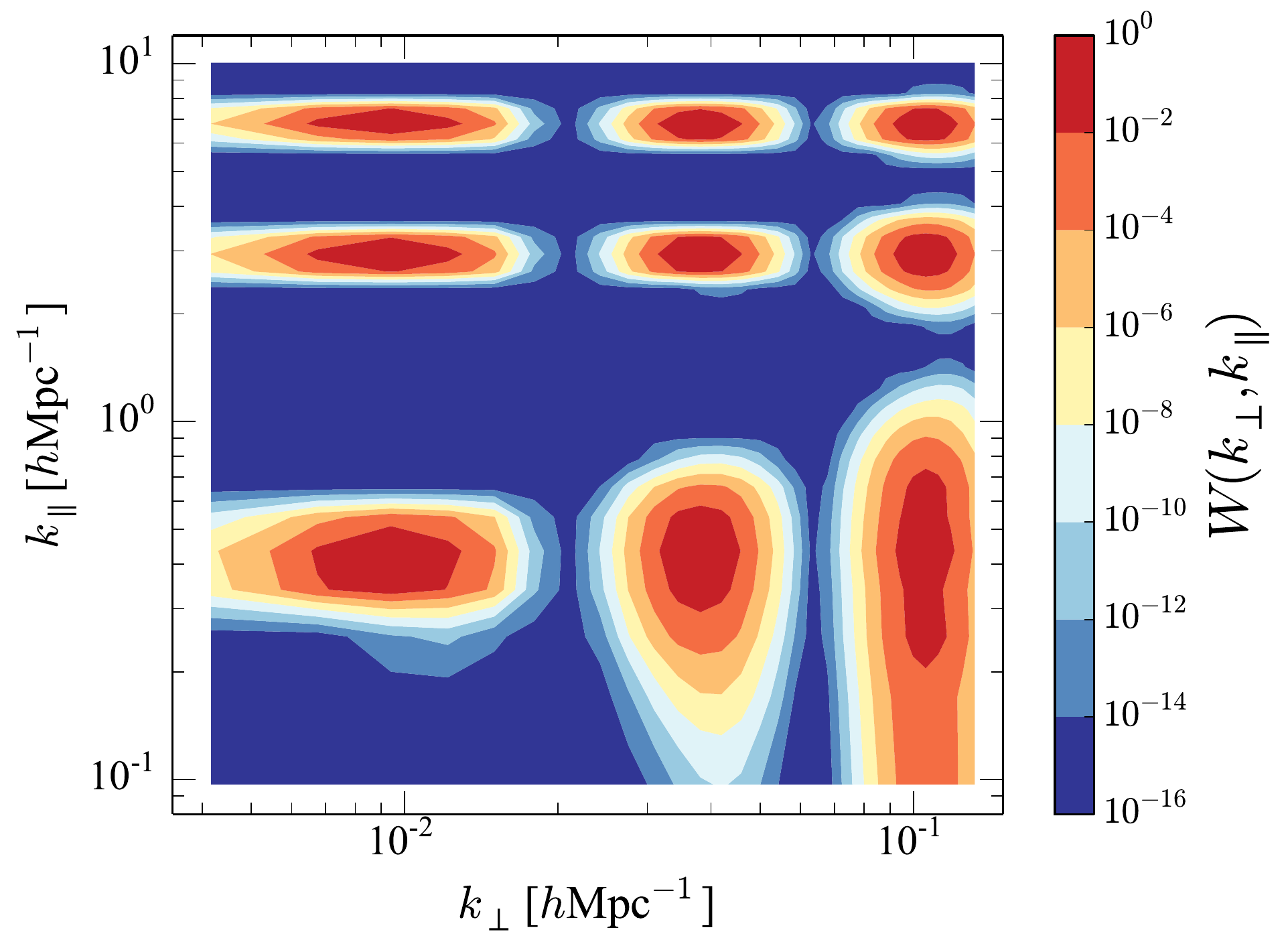}
	\caption{Sample window functions computed for the basic estimator.  From top left to bottom right, the windows are centered at $(k_\perp, k_\parallel) = (0.0094, 2.9)$, $(0.038, 2.9)$, $(0.11, 2.9)$, $(0.0094, 1.4)$, $(0.038, 1.4)$, $(0.11, 1.4)$, $(0.0094, 0.44)$, $(0.038, 0.44)$, $(0.11, 0.44)\,h\textrm{Mpc}^{-1}$.  Those centered at high $k_\perp$ and low $k_\parallel$ develop tails towards low $k_\parallel$, allowing foreground power to leak to higher $k_\parallel$.}
	\label{fig:basicEstWindowsSummary}
\end{figure}

\begin{figure}[!ht] 
	\centering 
	\includegraphics[width=.49\textwidth]{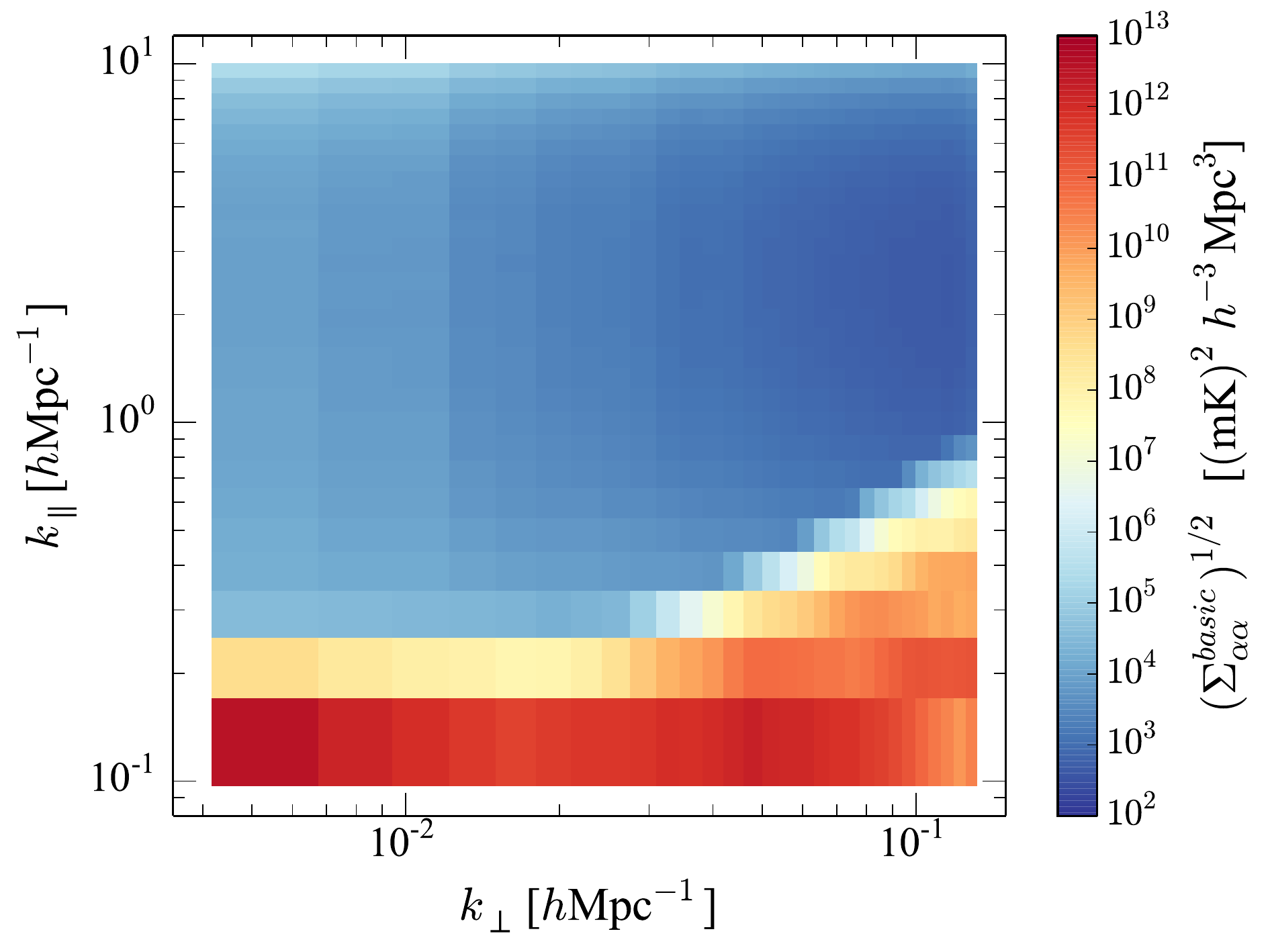}
	\caption{Power spectrum errors bars for our basic estimator, reproduced from Paper I.  The patterns are similar to those seen in the bias, with large errors driven by the foregrounds inside the wedge.  Errors drop significantly outside the wedge as they become dominated by the (much smaller) errors from thermal noise.}
	\label{fig:basicEstErrors}
\end{figure}

In Paper I, we examined a ``basic" estimator defined by the estimator matrix $\mathbf{E}_\alpha \propto \mathbf{N}^{-1} \mathbf{C}_{,\alpha} \mathbf{N}^{-1}$.  It was shown that in the limit of a very finely discretized $k_\perp k_\parallel$ grid, such an estimator is equivalent to one where visibilities are gridded with the Fourier footprint of the primary beam, followed by a squaring and binning of the resulting Fourier amplitudes.  Our basic estimator is therefore intended to be representative of many existing power spectrum pipelines \cite{Bernardi2013,Thyagarajan2013,Hazelton2013}.  Numerical computations gave the following conclusions:
\begin{enumerate}
\item[(1)] Window functions are typically quite narrow in extent outside high $k_\perp$ and low-to-moderate $k_\parallel$ regions (i.e., outside the wedge region).  As one encroaches upon the wedge, the window functions develop long tails towards low $k_\parallel$.  This causes spectrally smooth (low $k_\parallel$) foregrounds to be scattered to higher $k_\parallel$ modes in a $k_\perp$-dependent way, resulting in the wedge.  Some sample window functions from the basic estimator are shown in Figure \ref{fig:basicEstWindowsSummary}.
\item[(2)] The residual bias is shown in Figure \ref{fig:basicEstBias}.  This confirms the basic picture of strong foreground contamination inside the wedge, with a sharp drop-off as one moves away from the wedge and into the EoR window.
\item[(3)] Error bars are shown in Figure \ref{fig:basicEstErrors}.  The trends are similar to those seen in the bias: large errors exist within the wedge, but drop off sharply outside.  Note that whereas the bias in the EoR window is essentially zero, the errors plateau to a non-zero value.  This reflects the fact that cross-correlations remove noise bias, but not noise variance.
\item[(4)] Computing the full error covariance matrix $\boldsymbol \Sigma$, we found that off-diagonal elements (i.e., error correlations) were large.  In our basic estimator, Fourier modes within the wedge were seen to be highly correlated with each other, reducing the number of independent modes measured.  This suggests that sensitivity estimates for power spectrum measurements that use our basic estimator (or almost equivalently, by a simple squaring and binning following primary beam gridding on the Fourier plane) may be overly optimistic.
\end{enumerate}

These results from Paper I confirmed the basic picture of the foreground wedge and the EoR window in a statistically rigorous fashion, allowing a robust program of foreground avoidance where power spectrum measurements are made only outside the wedge.  As shown in Ref. \cite{Pober2014}, however, much-improved constraints on astrophysical parameters can be obtained if foregrounds can be mitigated sufficiently to enable measurements inside the wedge.  We will thus spend the rest of this paper examining various methods for reducing the extent of the foreground wedge, or equivalently, enlarging the EoR window.

\section{Direct subtraction of foregrounds}
\label{sec:Direct}
The most conceptually straightforward way to enlarge the EoR window is to directly subtract foregrounds from the data.  Trivially, if a perfect foreground model (containing all necessary spatial and spectral information) were to exist, one would be able to propagate this model through a measurement equation to predict measured visibilities.  These visibilites could then be subtracted from the data prior to power spectrum estimation, eliminating the foreground contribution $\mathbf{C}_\textrm{fg}$ from the total covariance $\mathbf{C}$.  In practice, foreground models are not perfect, and some residual foreground covariance will remain.  It is then the foreground \emph{uncertainty}, not the foregrounds themselves that are the biggest problem in power spectrum estimation, a point that was emphasized in Ref. \cite{Trott2012}.

For direct foreground subtraction to be effective, however, it is important to avoid inaccuracies, even though uncertainties are tolerable.  In other words, large error bars are permitted, but biases should be avoided at all costs.  This is because the foregrounds are typically much brighter than the cosmological signal, and thus small mis-subtractions of foregrounds may result in substantial signal loss.  Ideally, such signal loss should be quantified and adjusted for in one's final results, as was done in Refs. \cite{Masui2013,Paciga2013,Switzer2013}.  Additionally, mis-subtracted foregrounds may result in residual biases in the power spectrum, which are equally problematic.  In contrast, the procedures that we follow in this paper are much more conservative ones that use only the statistical properties of the foregrounds.  Ultimately, a combination of direct subtraction and statistical mitigation might be used, although a detailed study of the associated trade-off between the aggressiveness of the foreground subtraction and its robustness is necessary.

\section{Enlarging the EoR window with better estimators}
\label{sec:Enlarging}
In this section, we discuss how different estimators can be used to enlarge the EoR window.  This is motivated by the fact that were it not for the foregrounds, the ratio of the cosmological signal to thermal noise would be largest at low $k$.  As was pointed out in Ref. \cite{Pober2014}, working at low $k$ can result in large increases in detection significance, and ultimately to corresponding decreases in the error bars of measured model parameters.

In what follows, we will consider estimator matrices $\mathbf{E}_\alpha$ that fall into two distinct categories.  The first category consists of estimators that are separable, so that the estimator matrix may be written as
\begin{equation}
\label{eq:SepEst}
\mathbf{E}_\alpha \equiv \mathbf{e}_\alpha \mathbf{e}_\alpha^\dagger,
\end{equation}
i.e., as the outer product of some vector $\mathbf{e}_\alpha$ with itself.  With such estimators, the estimated bandpower $\widehat{p}_\alpha$ for the $\alpha$th cell in three-dimensional Fourier $\mathbf{k}$ space  takes the form
\begin{equation}
\label{eq:PreBinSepEst}
\widehat{p}_\alpha = \mathbf{x}^\dagger \mathbf{e}_\alpha \mathbf{e}_\alpha^\dagger \mathbf{x} = | \mathbf{e}_\alpha^\dagger \mathbf{x} |^2 \equiv | z_\alpha |^2.
\end{equation}
Separable estimators are therefore ones where linear combinations of the data are formed to measure Fourier amplitudes $z_\alpha$, which are then squared to give power spectra.  These power spectra can then be binned together according to the symmetry of the problem, an example being the cylindrical binning where all bandpowers falling inside an annular region of volume $V_A$ within a narrow range in $k_\perp$ and $k_\parallel$ are summed together:
\begin{equation}
\widehat{p}_A \equiv \frac{1}{V_A} \sum_{\alpha \in A} w_\alpha \widehat{p}_\alpha,
\end{equation}
where $w_\alpha$ is the weight given to the $\alpha$th bandpower.  While this last binning step means that the overall estimator of the binned power spectrum is not separable, we will continue to classify estimators as separable if they are separable prior to binning.

In contrast, non-separable estimators cannot be written in the form specified by Eq. \eqref{eq:SepEst}, even before binning.  Each bandpower is a linear combination not of the data vector elements themselves, but of product pairs of the data:
\begin{equation}
\label{eq:NonSepEst}
\widehat{p}_\alpha =  \mathbf{x}^\dagger \mathbf{E}_\alpha \mathbf{x} = \sum_{ij} (\mathbf{E}_\alpha)_{ij} \mathbf{x}_i \mathbf{x}_j.
\end{equation}
Compared to separable estimators, the non-separable estimators provide additional degrees of freedom in determining how the measured data ought to be weighted.  In general, then, the optimal power spectrum estimator will be of a non-separable form.  However, separable estimators are computationally easier to implement, and have therefore been used frequently in $21\,\textrm{cm}$ power spectrum estimation \cite{Bernardi2013,Pober2013b,Parsons2013}.  More specifically, suppose one wishes to estimate $N_\textrm{bands}$ bandpowers using a data vector that is of length $N_\textrm{meas} = N_\textrm{bl} N_\nu$ (where $N_\textrm{bl}$ is the number of baselines in our array, and $N_\nu$ is the number of frequency channels).  Separable estimators essentially involve evaluating a single dot product for each band, and therefore scale as $\mathcal{O}(N_\textrm{meas} N_\textrm{bands}) $.  In contrast, non-separable estimators scale as $\mathcal{O}(N_\textrm{meas}^2 N_\textrm{bands}) $ because a vector-matrix multiplication is involved.  Furthermore, these scalings assume that $\mathbf{E}_\alpha$ and $\mathbf{e}_\alpha$ are known ahead of time.  Including the time needed to compute these quantities often makes the difference in computational cost even more extreme.  For example, the optimal separable estimator that we derive in Section \ref{sec:FKP} makes use of a set of $\mathbf{e}_\alpha$ vectors that can be computed in $\mathcal{O}(N_\textrm{meas} N_\textrm{bands}) $ time.  On the other hand, constructing $\mathbf{E}_\alpha$ for the optimal non-separable estimator described in Section \ref{sec:BetterEst} requires evaluating $\mathbf{C}^{-1}$ in addition to the multiplication of several $N_\textrm{meas} \times N_\textrm{meas}$ matrices, giving a scaling of $\mathcal{O}(N_\textrm{meas}^3 N_\textrm{bands})$.  Additional assumptions regarding foreground and survey properties may potentially allow improvements to be made to this last scaling \cite{Dillon2013}, but it remains an open questions as to whether or not these assumptions are warranted at detection-level sensitivities and beyond.

\subsection{Separable estimators: minimizing the variance with the FKP approximation}
\label{sec:FKP}
In Paper I, we showed that in the limit of infinitely fine $k_\perp$ and $k_\parallel$ bins, the basic estimator that we summarized in Section \ref{sec:PaperI} is equivalent to a separable estimator where the Fourier amplitudes are estimated using a least-squares estimator, and then squared to form the power spectrum.  However, even though a correctly-normalized least-squares estimator is a provably optimal (minimum variance) way to form a map of Fourier amplitudes of the sky \cite{Tegmark1997a}, it does not follow that simply squaring those amplitudes is the optimal way to estimate the power spectrum.  In this section, we derive an improved separable power spectrum estimator by solving a constrained minimization problem: if computational limitations dictate that we form our power spectrum using Eq. \eqref{eq:SepEst} rather than Eq. \eqref{eq:NonSepEst}, what choice of $\mathbf{e}_\alpha$ gives the smallest error bars? The result will essentially be the Feldman-Kaiser-Peacock (FKP) approximation that is commonly used in galaxy surveys \cite{Feldman1994, Tegmark1998}, but applied to radio interferometers.  Our derivation will draw heavily from that presented in Ref. \cite{Tegmark1998} for galaxy surveys.

We begin with the visibility $V(\mathbf{b},\nu)$ measured by a single baseline $\mathbf{b}$ of an interferometer at frequency $\nu$:
\begin{equation}
\label{eq:MeasurementEqn}
V(\mathbf{b},\nu) = \int I(\boldsymbol \theta, \nu) A \left(\frac{\boldsymbol \theta}{\theta_0}, \nu \right) \exp\left(-i 2 \pi \frac{\nu}{c} \mathbf{b} \cdot \boldsymbol \theta \right) d^2 \boldsymbol \theta,
\end{equation}
where $I(\boldsymbol \theta, \nu)$ is the sky temperature, $A (\boldsymbol \theta / \theta_0, \nu)$ is the primary beam, and $\theta_0$ is some fiducial beam width.  A visibility is essentially a measure of a Fourier mode of the primary beam-weighted sky, with the wavenumber of the mode given by $\mathbf{u} = \nu \mathbf{b}/c$.  Now, suppose we were to ignore the inherent chromaticity of the interferometer, replacing $\nu$ in the exponent with $\nu_0$.  From the results of Paper I and Ref. \cite{Parsons2012b}, we know that this is tantamount to working only with short baselines, or being far away from the wedge.  With such a substitution, each visibility probes the same spatial Fourier mode across all frequencies.  An interferometer therefore measures sky modes in a hybrid space where the angular directions are cast in a Fourier basis, while the line-of-sight direction remains in a frequency basis.

To estimate Fourier amplitudes in full three-dimensional Fourier space, one can imagine first reconstructing a dirty image from the visibilities, weighting the result, and then Fourier transforming along both the angular and spectral directions to re-enter Fourier space:
\begin{equation}
\label{eq:zalpha}
z_\alpha = \int d^3\mathbf{r}\, e^{-i 2\pi \mathbf{q}_\alpha \cdot \mathbf{r}} \chi (\mathbf{r}) \sum_i e^{i 2 \pi \mathbf{u}_i \cdot \boldsymbol \theta} V(\mathbf{b}_i = \mathbf{u}_i c / \nu_0, \nu),
\end{equation}
where we have defined $\mathbf{q} \equiv (\mathbf{u}, \eta)$ as the Fourier dual to $\mathbf{r} \equiv (\boldsymbol \theta, \nu)$, and $\chi(\mathbf{r})$ is a weighting function.\footnote{In this section, we adopt a Fourier convention where the Fourier transform $\widetilde{f} (\mathbf{q})$ is given by $\int d^3 \mathbf{r} e^{-i 2\pi \mathbf{q} \cdot \mathbf{r}} f(\mathbf{r})$ and the inverse transform by $f(\mathbf{r}) = \int d^3 \mathbf{q} e^{i 2\pi \mathbf{q} \cdot \mathbf{r}} \widetilde{f} (\mathbf{q})$.  This convention is convenient because of its similarities to the definition of a visibility, and differs from the cosmological convention in two ways.  First, our real-space position vector $\mathbf{r}$ is comprised of two angular coordinates and a frequency, so cylindrical power spectra formed under our convention inhabit $\mathbf{u}\eta$ space rather than $\mathbf{k}$ space.  Their conventions also differ in their placement of factors of $2\pi$; the cosmological convention has no such factors in the exponents, and instead has a $(2\pi)^3$ in the denominator of the inverse transform.  From Section \ref{sec:BetterEst} onwards, we will express our results using the cosmological convention to allow an easier comparison to other works in the literature.  For explicit expressions for converting between the two conventions, see Appendix A of Paper I.}  Note that since we have yet to specify the form of $\chi$, our expression for $z_\alpha$ is flexible enough to describe many common prescriptions for imaging.  For example, since a multiplicative weighting in the image domain is equivalent to a convolution in $uv$ space, this includes  the prescription suggested in Ref. \cite{Tegmark1997a,Morales2008}, where the visibilities are re-gridded with a $uv$ space primary beam kernel.  Our goal in this section will be to find a form for $\chi$ that minimizes the variance of our separable power spectrum estimator $\widehat{p}_\alpha =  | z_\alpha |^2$.

For analytical tractability, it is convenient to replace the sum in Eq. \eqref{eq:zalpha} with an integral over a  continuous distribution $ \tilde{\rho} (\mathbf{u})$ of baselines on the $uv$ plane, i.e.
\begin{equation}
\sum_i  ( \dots ) \rightarrow \int d^2 \mathbf{u}_i \, \tilde{\rho} (\mathbf{u}_i)  ( \dots ),
\end{equation}
where $ \tilde{\rho} (\mathbf{u})$ is normalized such that $\int d^2 \mathbf{u} \, \tilde{\rho} (\mathbf{u}) $ equals the total number of baselines used in the analysis.  Making this substitution and inserting Eq. \eqref{eq:MeasurementEqn} into Eq. \eqref{eq:zalpha} gives
\begin{equation}
z_\alpha = \int d^3\mathbf{r}d^2 \boldsymbol \theta^\prime \, e^{-i 2\pi \mathbf{q}_\alpha \cdot \mathbf{r}} \chi (\mathbf{r})   I(\boldsymbol \theta^\prime, \nu) A \left(\frac{\boldsymbol \theta^\prime}{\theta_0}, \nu \right) 
\rho( \boldsymbol \theta - \boldsymbol \theta^\prime).
\end{equation}
The function $\rho$ is the synthesized beam of the interferometer array, and is given by the inverse Fourier transform of $\tilde{\rho}$ over the spatial directions.
Now, the primary beam is typically much smoother than the sky signal, and varies on a characteristic length scale that is much wider than that of the synthesized beam.  We may therefore factor the primary beam out of the $\boldsymbol \theta^\prime$ integral, and applying Parseval's theorem to the result gives
\begin{eqnarray}
z_\alpha &=& \int d^3 \mathbf{r} \,d^3 \mathbf{q}\, e^{-i 2 \pi (\mathbf{q}_\alpha - \mathbf{q}) \cdot \mathbf{r}} \chi (\mathbf{r}) A(\mathbf{r}) \tilde{\rho} (\mathbf{u}) \widetilde{I} (\mathbf{q}) \nonumber \\
&=&\int d^3 \mathbf{q}\, \widetilde{\psi}(\mathbf{q}_\alpha - \mathbf{q}) \tilde{\rho} (\mathbf{u}) \widetilde{I} (\mathbf{q}) ,
\end{eqnarray}
where $\widetilde{I}$ is the Fourier transform of the sky temperature, and $\widetilde{\psi}$ is the Fourier transform of the newly-defined $\psi (\mathbf{r}) \equiv \chi(\mathbf{r}) A(\mathbf{r})$.  The signal covariance matrix is then given by
\begin{eqnarray}
\label{eq:NotYetFactoredSab}
\mathbf{S}_{\alpha \beta} &\equiv& \langle z_\alpha z_\beta^* \rangle \nonumber \\
&=& \int d^3 \mathbf{q} \,\tilde{\rho} (\mathbf{u})^2 P(\mathbf{q}) \widetilde{\psi}(\mathbf{q}_\alpha - \mathbf{q}) \widetilde{\psi}^*(\mathbf{q}_\beta - \mathbf{q}), \qquad
\end{eqnarray}
where we have used the definition of the power spectrum, $\langle \widetilde{I}(\mathbf{q}) \widetilde{I}(\mathbf{q^\prime})^* \rangle \equiv P(\mathbf{q}) \delta(\mathbf{q} - \mathbf{q}^\prime)$.  In the literature, it is often assumed that the power spectrum is smooth and slowly varying.  In contrast, as long as $\chi (\mathbf{r})$ is not taken to be too narrow, a relatively broad primary beam $A(\mathbf{r})$ will ensure that $\widetilde{\psi}$ is sharply peaked.  We may therefore factor $P(\mathbf{q})$ out of the integral \cite{McQuinn2006,Parsons2012a}.  Assuming that $\tilde{\rho} (\mathbf{u})$ is also smooth, this gives
\begin{eqnarray}
\mathbf{S}_{\alpha \beta} &\equiv& \tilde{\rho}^2 P \int d^3 \mathbf{q} \,\widetilde{\psi}(\mathbf{q}_\alpha - \mathbf{q}) \widetilde{\psi}^*(\mathbf{q}_\beta - \mathbf{q}) \nonumber \\
&=& \tilde{\rho}^2 P \int d^3 \mathbf{r} \, \chi^2 (\mathbf{r}) A^2 (\mathbf{r}) e^{-i 2 \pi (\mathbf{q}_\alpha - \mathbf{q}_\beta) \cdot \mathbf{r}},
\end{eqnarray}
where we have defined $\tilde{\rho} \equiv \tilde{\rho} [(\mathbf{u}_\alpha + \mathbf{u}_\beta)/2]$, $P\equiv P [(\mathbf{q}_\alpha + \mathbf{q}_\beta)/2]$, and in the last equality we used Parseval's theorem.  The expected value of our estimator is given by the diagonal elements of this matrix, i.e.
\begin{equation}
\langle \widehat{p}_\alpha \rangle =  \langle | z_\alpha |^2 \rangle = \mathbf{S}_{\alpha \alpha} = \tilde{\rho}^2 P \int d^3 \mathbf{r} \, \chi^2 (\mathbf{r}) A^2 (\mathbf{r}) .
\end{equation}
With this being proportional to $P$, we see that we can arrange for our estimator to have no multiplicative bias (i.e., for it to have the property that $\langle \widehat{p}_\alpha \rangle= P$) if $\chi(\mathbf{r})$ is correctly normalized.  The optimization that we perform below will determine the form of $\chi(\mathbf{r})$ only to an overall constant, which affords us the freedom to do so.

To compute the error bars on this mean power spectrum estimate, we also require the instrumental noise contribution to the total covariance.  The instrumental noise enters as an additive contribution $n(\mathbf{b}_i,\nu)$ to the visibility.  Inserting $n(\mathbf{b}_i,\nu)$ in the place of $V(\mathbf{b}_i, \nu)$ in Eq. \eqref{eq:zalpha} gives the noise contribution to the Fourier amplitude $n_\alpha$.  Forming the noise covariance $\mathbf{N}_{\alpha \beta} \equiv \langle n_\alpha n_\beta^* \rangle$ then yields
\begin{eqnarray}
\mathbf{N}_{\alpha \beta} = \int d^3 \mathbf{r} d^3 \mathbf{r}^\prime e^{-i 2\pi \mathbf{q}_\alpha \cdot \mathbf{r}} e^{i 2 \pi \mathbf{q}_\beta \cdot \mathbf{r}^\prime} \chi( \mathbf{r} ) \chi( \mathbf{r}^\prime) \times \nonumber \\
\sum_{ij} e^{i 2 \pi \frac{\nu}{c} \mathbf{b}_i \cdot \boldsymbol \theta} e^{-i 2 \pi \frac{\nu^\prime}{c} \mathbf{b}_j \cdot \boldsymbol \theta^\prime} \langle n(\mathbf{b}_i,\nu) n(\mathbf{b}_j,\nu^\prime)^* \rangle.
\end{eqnarray}
To simplify this expression, we assume that the instrumental noise is uncorrelated between baselines and frequency channels, so that
\begin{equation}
 \langle n(\mathbf{b}_i,\nu) n(\mathbf{b}_j,\nu^\prime)^* \rangle = \sigma^2 (\nu) B_\textrm{chan} \delta(\nu- \nu^\prime) \delta_{ij},
\end{equation}
where $\sigma^2$ is the noise variance in a single frequency channel, and the factor of $B_\textrm{chan}$ (the frequency channel width) accounts for the fact that instrumental noise is correlated \emph{within} a frequency channel.  Inserting this into our noise covariance gives
\begin{eqnarray}
\mathbf{N}_{\alpha \beta}  =  && \int d^2 \boldsymbol \theta d^2 \boldsymbol \theta^\prime d\nu e^{-i 2\pi \mathbf{u}_\alpha \cdot \boldsymbol \theta}  e^{i 2\pi \mathbf{u}_\beta \cdot \boldsymbol \theta^\prime} e^{-i 2\pi (\eta_\alpha - \eta_\beta) \nu} \times \nonumber \\
&& \chi(\boldsymbol \theta, \nu) \chi(\boldsymbol \theta^\prime, \nu) \sigma^2 (\nu) B_\textrm{chan} \sum_i e^{i 2\pi \frac{\nu}{c} (\mathbf{b}_i \cdot \boldsymbol \theta - \mathbf{b}_j \cdot \boldsymbol \theta^\prime)} \nonumber \\
= && \sum_i \int d\nu\, \sigma^2 (\nu) B_\textrm{chan} e^{-i 2\pi (\eta_\alpha - \eta_\beta) \nu} \times \nonumber \\
&& \overline{\chi} \left(\mathbf{u}_\alpha - \frac{\nu \mathbf{b}_i}{c}, \nu \right) \overline{\chi}^* \left(\mathbf{u}_\beta - \frac{\nu \mathbf{b}_i}{c}, \nu \right),
\end{eqnarray}
where $\overline{\chi}$ is the Fourier transform of $\chi$ over the spatial directions \emph{only}.  We can proceed further by once again neglecting the chromaticity of the interferometer (by letting $\nu \mathbf{b}_i / c \approx \nu_0 \mathbf{b}_i / c \equiv \mathbf{u}_i$), and by replacing the discrete sum over baselines with a continuous distribution over $\mathbf{u}_i$:
\begin{eqnarray}
\mathbf{N}_{\alpha \beta} = \int d\nu  &&d^2\mathbf{u}_i e^{-i 2\pi (\eta_\alpha - \eta_\beta) \nu} \sigma^2 (\nu) B_\textrm{chan}\tilde{\rho}(\mathbf{u}_i) \times \nonumber \\
&&  \overline{\chi} \left(\mathbf{u}_\alpha -  \mathbf{u}_i, \nu \right) \overline{\chi}^* \left(\mathbf{u}_\beta - \mathbf{u}_i, \nu \right)
\end{eqnarray}
As before, we assume that $\chi(\mathbf{r})$ is smooth and broad (this can self-consistently checked later), which means that $\overline{\chi}$ will be narrowly peaked.  The $uv$ coverage density $\tilde{\rho}$ may therefore be factored out of the integral, and invoking Parseval's theorem in the spatial directions yields
\begin{equation}
\mathbf{N}_{\alpha \beta} = \tilde{\rho} \int d^3 \mathbf{r} \, \sigma^2 (\nu) B_\textrm{chan} \chi^2 (\mathbf{r}) e^{-i 2\pi (\mathbf{q}_\alpha - \mathbf{q}_\beta)}.
\end{equation}
We therefore have a total covariance $\mathbf{C} \equiv \mathbf{S} + \mathbf{N}$ given by
\begin{equation}
\label{eq:Ctot}
\mathbf{C}_{\alpha \beta} =\int d^3 \mathbf{r} \,\  \chi^2 (\mathbf{r}) \left[ \tilde{\rho} \sigma^2 B_\textrm{chan}  + \tilde{\rho}^2 P A^2 (\mathbf{r}) \right] e^{-i 2 \pi (\mathbf{q}_\alpha - \mathbf{q}_\beta) \cdot \mathbf{r}}.
\end{equation}
As long as $A(\mathbf{r})$ and $\chi(\mathbf{r})$ are smooth and broad, this shows that the $\alpha$th and $\beta$th Fourier cells will be correlated only if $|\mathbf{q}_\alpha - \mathbf{q}_\beta|$ is small.  In the $\mathbf{u}$ direction, this scale is set by the Fourier transform of the primary beam, and in the $\eta$ direction, the inverse of the observation bandwidth $B_\textrm{band}$.  In binning squared Fourier amplitudes to form the cylindrical power spectrum, it therefore makes sense to bin in annuli whose characteristic dimensions are larger than these scales, so that different annuli are uncorrelated.  Assuming that rotation synthesis yields a $uv$ coverage that is approximately azimuthally symmetric, modes falling within the same annulus can be summed with equal weights, and the resulting variance $\Delta p_A$ in our binned estimator $\widehat{p}_A$ is
\begin{equation}
\label{eq:Binning}
(\Delta p_A)^2 = \frac{2}{V_A^2} \int \,d^3 \mathbf{q}_\alpha d^3 \mathbf{q}_\beta | \mathbf{C}_{\alpha \beta} |^2  .
\end{equation}
These integrals, which both extend over $V_A$, can be evaluated by making the substitutions $\mathbf{q}_- \equiv \mathbf{q}_\alpha - \mathbf{q}_\beta$ and $\mathbf{q}_+ \equiv (\mathbf{q}_\alpha + \mathbf{q}_\beta)/2$.  With such a substitution, the $\mathbf{q}_+$ integral becomes trivial and simply gives a factor of $V_A$.  We then have
\begin{eqnarray}
(\Delta p_A)^2 = \frac{2}{V_A} \int d^3\mathbf{r} d^3\mathbf{r}^\prime d^3 \mathbf{q} e^{-i 2\pi \mathbf{q}_- \cdot (\mathbf{r} - \mathbf{r}^\prime)}  \chi^2 (\mathbf{r}) \chi^2 (\mathbf{r}^\prime) \times \nonumber \\
 \left[ \tilde{\rho} \sigma^2 B_\textrm{chan}  + \tilde{\rho}^2 P A^2 (\mathbf{r}) \right] \left[ \tilde{\rho} \sigma^2 B_\textrm{chan}  + \tilde{\rho}^2 P A^2 (\mathbf{r}^\prime) \right] \qquad
\end{eqnarray}
For the $\mathbf{q}_-$ integral, we can extend the bounds of the integral to $\pm \infty$, since $|\mathbf{C}_{\alpha \beta}|^2 \approx 0$ for large $|\mathbf{q}_-|$ anyway.  The complex exponentials from Eq. \eqref{eq:Ctot} then integrate to give $\delta(\mathbf{r} - \mathbf{r}^\prime)$, and thus
\begin{equation}
(\Delta p_A)^2 = \frac{2}{V_A} \int d^3 \mathbf{r} \, \chi^4 (\mathbf{r}) \left[ \tilde{\rho} \sigma^2(\nu) B_\textrm{chan}  + \tilde{\rho}^2 P A^2 (\mathbf{r}) \right]^2.
\end{equation}

To optimize our separable estimator, we seek a functional form for $\chi(\mathbf{r})$ that minimizes $(\Delta p_A)^2$, subject to the constraint that $\langle \widehat{p}_A \rangle \equiv (1/V_A) \sum_{\alpha \in A} \widehat{p}_\alpha$ be held constant.  This can be accomplished using a Lagrange multiplier $\lambda$, minimizing $(\Delta p_A)^2 - \lambda \langle \widehat{p}_A \rangle$.  The result is
\begin{equation}
\label{eq:optWeights}
\chi(\mathbf{r}) \propto \frac{\tilde{\rho}^2 P A(\mathbf{r}) }{\tilde{\rho}^2 P A^2(\mathbf{r}) +\tilde{\rho} \sigma^2(\nu) B_\textrm{chan} },
\end{equation}
and inserting this back into our equation for the variance gives a convenient expression for the resultant errors:
\begin{equation}
\label{eq:PercentError}
\frac{\Delta P(\mathbf{u},\eta)}{\langle \widehat{P}(\mathbf{u}, \eta) \rangle}\equiv \frac{\Delta p_A}{\langle  \widehat{p}_A \rangle} = \sqrt{\frac{2}{V_A V_\textrm{eff}}},
\end{equation}
with
\begin{equation}
\label{eq:Veff}
V_\textrm{eff} (\mathbf{u}, \eta) \equiv \int d^3 \mathbf{r} \frac{\tilde{\rho}^4 (\mathbf{u}) P^2 (\mathbf{u}, \eta) A^4 (\mathbf{r})}{[\tilde{\rho}^2 (\mathbf{u}) P (\mathbf{u}, \eta) A^2 (\mathbf{r}) + \tilde{\rho}(\mathbf{u}) \sigma^2(\nu) B_\textrm{chan}]^2},
\end{equation}
where we have restored the $\mathbf{u}$ and $\eta$ dependencies of $\tilde{\rho}$ and $P(\mathbf{u}, \eta)$.  This expression, which tends to $V_A$ in the high signal-to-noise limit, is the radio interferometer equivalent of the effective volume $V_\textrm{eff}$ quoted in galaxy surveys.  The weighting function $\chi (\mathbf{r})$ is the radio interferometric analog of the FKP weighting function.

To get some intuition for Eqs. \eqref{eq:PercentError} and \eqref{eq:Veff}, we make use of the fact that
\begin{equation}
\sigma^2 (\nu) = \frac{T^2_\textrm{sys}}{2 B_\textrm{band} t} \Omega_{pp}^2,
\end{equation}
where $\Omega_{pp} \equiv \int d^2 \boldsymbol \theta A^2 ( \boldsymbol \theta)$, $t$ is the integration time on the $\mathbf{u}$ mode that one is seeking to measure, and $T_\textrm{sys}$ is the system temperature.  In the noise-dominated regime, the first term in the denominator of $V_\textrm{eff}$ dominates, 
 and $V_\textrm{eff} \propto \tilde{\rho}^2(\mathbf{u}) P^2 B_\textrm{band}^2 t^2 / T_\textrm{sys}^4 B_\textrm{chan}^2 $ (assuming that $T_\textrm{sys}$ is constant over our observing band, for pedagogical reasons).  This gives
 \begin{equation}
 \Delta P(\mathbf{u}, \eta) \Bigg{|}_\textrm{noise dom.} \propto \frac{T^2_\textrm{sys}}{t \tilde{\rho}(\mathbf{u})},
 \end{equation}
 where we have made use of the unbiased property $\langle \widehat{P}(\mathbf{u}, \eta) \rangle = P(\mathbf{u}, \eta)$ to cancel out the factor of $P^2$ in $V_\textrm{eff}$.  The scalings in this noise-dominated regime are in agreement with noise-only error estimates in the literature \cite{Morales2005,Parsons2012a}, although our constant factors differ slightly due to the FKP weighting scheme.
 
 In the low noise or high signal regime, $V_\textrm{eff}$ is approximately equal to the survey volume $V_s$.  The power spectrum errors are then
 \begin{equation}
 \Delta P(\mathbf{u}, \eta) \Bigg{|}_\textrm{signal dom.} = \sqrt{\frac{2}{V_A V_s}} P(\mathbf{u}, \eta).
 \end{equation}
 The errors are seen to be proportional to the power spectrum itself, in what would normally be denoted the cosmic variance-limited regime.  However, consider what ``cosmic variance" means in this case.  Our measurement of the sky power spectrum includes not just the cosmological signal but also any residual foregrounds.  Unless foregrounds can be pre-subtracted to an extremely high precision, it is therefore likely that the low instrumental noise regime will be dominated by a ``cosmic variance" of foregrounds.  In other words, foreground uncertainty will dominate the error bars, a fact that is missed by noise-only sensitivity calculations.

The FKP formalism shown here naturally interpolates between the instrumental noise- and foreground uncertainty-limited regimes.  However, in order to derive the radio interferometer FKP weights $\chi(\mathbf{r})$, it was necessary to make a number of approximations.  These approximations essentially all originate from the same assumption: that one is working on small scales both spatially and spectrally.  To see this, consider for example the annular binning in Eq. \eqref{eq:Binning}.  There, it was necessary to choose $|\mathbf{u}| = u$ bin edges $[u_i, u_{i+1}]$ that satisfied $\theta_0^{-1} \ll |u_{i+1} - u_i | \ll u $, where $u$ was the spatial Fourier wavenumber being probed (and similarly for $\eta$).  The first inequality ensured that power estimates in different annuli were uncorrelated, while the second equality ensured that each annulus would probe a relatively narrow range of wavenumbers.  This only holds on small scales, where $u \gg \theta_0^{-1}$.  This assumption was also necessary when we factored the power spectrum out of Eq. \eqref{eq:NotYetFactoredSab}.

In the appropriate small-scale regime, the FKP approximation has been proven to be optimal, in the sense that it delivers the smallest error bars amongst all possible estimators,\footnote{This fact is typically proven in the context of galaxy surveys, but the mathematics can be easily adapted for radio interferometry, as we have done for the derivation of the FKP approximation itself.} separable or not \cite{Tegmark1998}.  With this in mind, consider again the limit of a thermal noise-limited experiment.   If thermal noise is the limiting factor, the second term in the denominator of Eq. \eqref{eq:optWeights} dominates $\chi(\mathbf{r})$, and we essentially have $\chi(\mathbf{r}) \propto A(\mathbf{r})$.  Since the multiplicative application of $A (\mathbf{r})$ in the image domain is equivalent to a convolution of its spatial Fourier transform on the $uv$ plane, we see that in the high-noise regime, the FKP approach is equivalent to the commonly-used recipe of first gridding the visibility measurements on the $uv$ plane with the Fourier-space primary beam function, then Fourier transforming along the frequency axis, and finally squaring and binning to find the power spectrum \cite{Bernardi2013,Thyagarajan2013,Hazelton2013}.  And since the FKP method is optimal on small scales, we should expect that at high $u$ and high $\eta$ (which is safely outside the wedge), excellent results should be attainable with either this recipe, or the closely-related basic estimator discussed in Paper I and Section \ref{sec:PaperI}.  In Section \ref{sec:BetterEst}, we shall find that this is indeed the case, where the basic estimator performs very well when safely within the confines of the EoR window.

Unfortunately, the oft-used recipe is unlikely to help as we attempt to push into the wedge region.  First, as one begins to move to lower $k_\perp$ and $k_\parallel$ values, the sky-signal-to-noise ratio increases.  The high-noise approximation for $\chi(\mathbf{r})$ therefore ceases to apply, and the full expression for the FKP weights ought to be applied instead.  Put another way, it is necessary to account for the fact that at lower $k$, the errors are dominated by foreground uncertainty, and not by thermal noise.  Having said this, even a full application of the FKP weights may be insufficient, because the push to lower $k_\parallel$ at high values of $k_\perp$ (necessary for enlarging the EoR window) explicitly violates a number of the assumptions necessary for the FKP approximation to be optimal.  For example, we violate the condition that we are working on small scales, as well as the short-baseline assumption in our derivation.  In addition, the edge of the wedge is (by definition) where one transitions from being relatively foreground-free to being foreground dominated.  The power spectrum must therefore evolve rapidly and be insufficiently slowly-varying to justify factoring it out of the integral in Eq. \ref{eq:NotYetFactoredSab}.  Finally, at high $k_\perp$ and low $k_\parallel$ (i.e. in the wedge), we saw in Paper I that errors were correlated over a much greater extent than just $\Delta u \sim \theta_0^{-1}$ and $\Delta \eta \sim B_\textrm{band}$, violating the assumptions we made when binning together power estimates.

In this section, we considered separable power spectrum estimators, and were able to derive a set of weights under the FKP approximation that were provably optimal in the small-scale limit.  Encouragingly, in the high-noise regime (suitable for current-generation experiments) this coincided with the basic power spectrum methods seen in the literature \cite{Bernardi2013,Thyagarajan2013,Hazelton2013}, which are also equivalent to the basic estimator of Section \ref{sec:PaperI} if the Fourier bins are taken to be extremely fine (in which case the estimator becomes separable---see Paper I for a proof of the equivalence between our basic estimator and existing methods).  However, the assumption of having foreground uncertainties be subdominant to thermal noise errors must necessarily be violated if one wishes to enlarge the EoR window by working within the foreground wedge.  Moreover, our derivation of the FKP approximation required us to make the assumption of short baselines, which is in conflict with the small-scale limit.  Put together, all these (sometimes conflicting) assumptions make it likely that enlarging the EoR window will require a more general, non-separable estimator.

%
%

\subsection{Non-separable estimators: minimizing the variance with optimal quadratic estimators}
\label{sec:BetterEst}

In Refs. \cite{Tegmark1997b,Bond1998}, it was shown that the optimal, minimum variance power spectrum estimator is given by\footnote{This can be proved using the Fisher matrix formalism, which we discuss briefly in Appendix \ref{appendix:Fisher}.}
\begin{equation}
\label{eq:OptEalpha}
\mathbf{E}_\alpha \propto \mathbf{C}^{-1} \mathbf{C}_{,\alpha} \mathbf{C}^{-1}.
\end{equation}
This is similar to the basic estimator considered in Paper I, except the data are weighted not by the inverse of the (diagonal) instrumental noise covariance, but instead by the full inverse covariance matrix $\mathbf{C}^{-1}$.  Doing so guarantees the smallest error bars possible, and as we shall see later in this section, the result is a slightly enlarged EoR window.

Because the total covariance $\mathbf{C}$ includes the foreground and cosmological signal covariances in addition to the instrumental noise covariance, off-diagonal elements must in principle be included when modeling the matrix.  The application of $\mathbf{C}^{-1}$ therefore not only downweights heavily contaminated modes based on the relative amplitudes of various modes, but also takes advantage of correlation information to deliver an optimal power spectrum estimate.  An approximate form of Eq. \eqref{eq:OptEalpha} that makes use of the former but not the latter is given by
\begin{equation}
\label{eq:CdiagEalpha}
\mathbf{E}_\alpha \propto \mathbf{C}^{-1}_\textrm{diag} \mathbf{C}_{,\alpha} \mathbf{C}^{-1}_\textrm{diag},
\end{equation}
where $\mathbf{C}^{-1}_\textrm{diag}$ is a diagonal approximation to the full inverse covariance matrix with elements $(\mathbf{C}_\textrm{diag}^{-1})_{ij} \equiv (\mathbf{C}_{ii})^{-1} \delta_{ij}$.  With off-diagonal elements omitted, the effect of $\mathbf{C}^{-1}_\textrm{diag}$ is to downweight parts of the data vector without using correlation information.  In what follows, we will compare both the optimal estimator and its approximate form to the basic estimator of Paper I and Section \ref{sec:PaperI}.

We stress, however, that as currently written, Eq. \eqref{eq:CdiagEalpha} is not well-defined.  This is because the diagonal of a matrix is a basis-specific quantity, and we have yet to specify the basis that we are working in.  It is therefore meaningless to ask whether it is a good approximation to neglect correlation information, for data that are correlated in one basis may be uncorrelated in another.  For instance, any matrix is (by construction) diagonal in its own eigenbasis.  However, working in the eigenbasis of $\mathbf{C}$ negates the power of an approximate treatment, for in general the switch into the eigenbasis is just as computationally difficult as computing the inverse.  The appropriate question to consider, then, is whether or not there exists a basis that is not only easy to switch to, but also has the property that correlation information can be neglected without having a significant impact on the final power spectrum error bars.

As a guess for such a basis, consider one where the input visibilities are grouped baseline-by-baseline, with each baseline's visibilities represented not by a frequency spectrum, but by a delay spectrum.  The delay spectrum $\widetilde{V} (\mathbf{b}, \tau)$ of a baseline is related to the visibility frequency spectrum $V(\mathbf{b}, \nu)$ by a Fourier transform \cite{Parsons2012b}:
\begin{equation}
\label{eq:DelayTransDef}
\widetilde{V} (\mathbf{b}, \tau) = \int V(\mathbf{b}, \nu) \phi \left( \frac{\nu-\nu_0}{B_\textrm{band}} \right) e^{-i 2\pi \nu \tau},
\end{equation}
where $\phi$ is a tapering function, and $B_\textrm{band}$ is the total bandwidth of the observation, centered at frequency $\nu_0$.  After a delay transform, visibilities are a function of the delay $\tau$.  Since the observed frequency of a spectral line (the $21\,\textrm{cm}$ line in our case) maps to line-of-sight distance, $\tau$ may be used (after multiplying by some constants) as an approximation for $k_\parallel$.  Similarly, each baseline probes a certain interference pattern (with a characteristic spatial scale) on the sky.  Delay spectra can therefore be approximately mapped to specific Fourier amplitudes of the three-dimensional sky.  This correspondence is not exact because visibilities measured by a single baseline probe different spatial modes of the sky at different frequencies.  Indeed, inserting Eq. \eqref{eq:MeasurementEqn} into Eq. \eqref{eq:DelayTransDef} shows that because of this coupling between spatial and spectral modes, a delay spectrum does not constitute a ``true" Fourier transform of the three-dimensional sky, which requires that each dimension's Fourier transform be performed independently.  In the limit of short baselines, however, Ref. \cite{Parsons2012b} showed that delay modes are an excellent approximation for true Fourier modes.  In addition, in Paper I we showed that in a baseline-delay basis---where the data vector $\mathbf{x}$ is a serialized list of delay spectra over all baselines and delays---the covariance matrix is close to diagonal in the same limit.  This makes the baseline-delay basis an attractive one to use in Eq. \eqref{eq:CdiagEalpha}.

\begin{figure*}[!ht] 
	\centering 
	\includegraphics[width=1\textwidth]{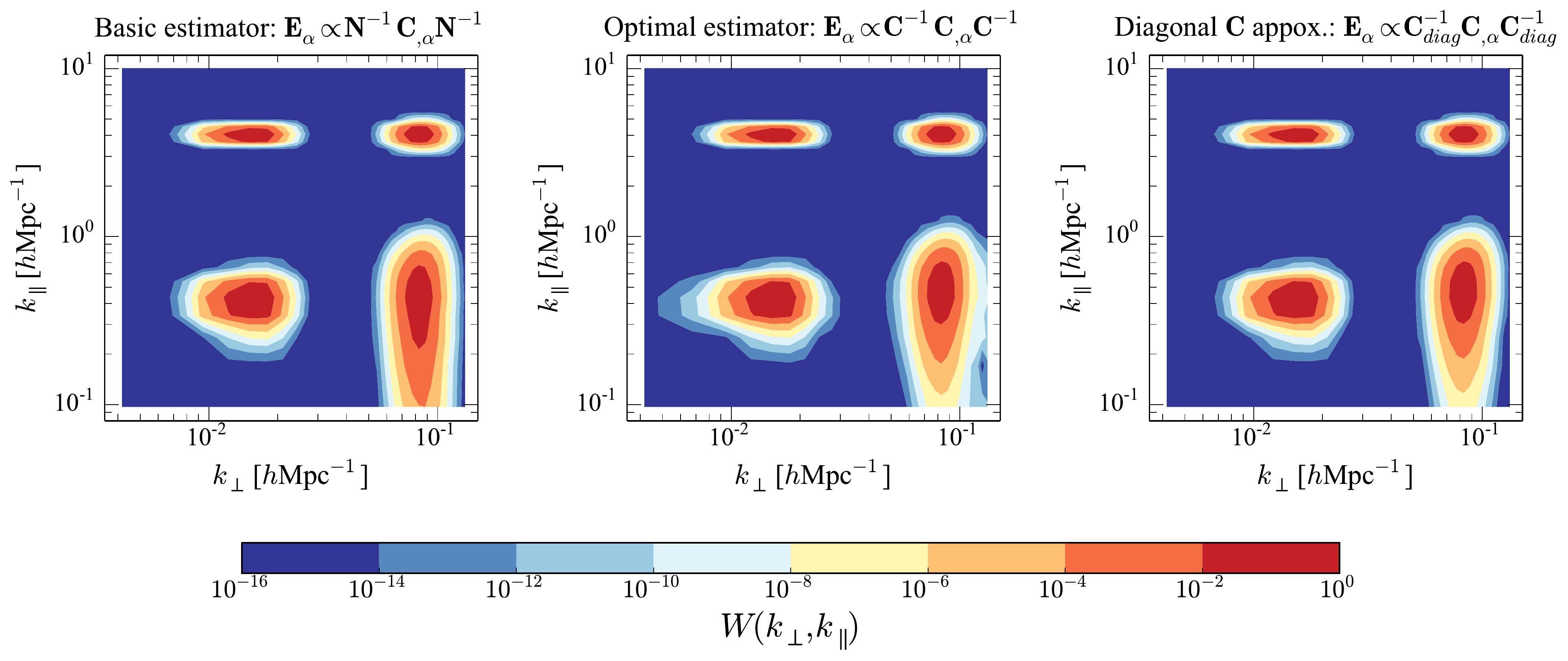}
	\caption{A comparison between the window functions of the basic estimator used in Paper I and Section \ref{sec:PaperI}, the optimal minimum variance estimator given by Eq. \eqref{eq:OptEalpha}, and the diagonal approximation estimator given by Eq. \eqref{eq:CdiagEalpha}.  Each plot shows four example window functions from one of the three estimators.  Going clockwise from top left to bottom left, the windows are centered at $(k_\perp, k_\parallel) = (0.015, 4.07)\,h\textrm{Mpc}^{-1}$, $(0.083, 4.07)\,h\textrm{Mpc}^{-1}$, $(0.083, 0.44)\,h\textrm{Mpc}^{-1}$, and $(0.015, 0.44)\,h\textrm{Mpc}^{-1}$.  Away from the wedge region, the window functions behave quite similarly, and therefore the error statistics should be comparable between all three estimators.  Close to the wedge, however, the window functions for both the optimal estimator and the diagonal approximation have shorter tails in the direction of the heavily foreground-contaminated low $k_\parallel$ region.}
	\label{fig:comparisonWindCollection}
\end{figure*} 

In Figure \ref{fig:comparisonWindCollection}, we show some example window functions for the three estimators we have discussed: the basic estimator from before, the optimal estimator, and the diagonal covariance approximation to optimal estimator.  Each plot shows four example windows for a particular estimator.  Three of the windows are chosen to be centered away from the the bottom right corner of the $k_\perp k_\parallel$ plane (i.e. away from the wedge region), while the fourth is chosen to sit roughly on the edge of the wedge.  

Consider first the example window functions centered on the edge of the wedge.  With all three estimators one sees the elongation in the $k_\parallel$ direction that was noted in Paper I and summarized in Section \ref{sec:PaperI}.  However, this elongation is less severe for the optimal and diagonal approximation estimators than it is for the basic estimator.  The estimators that we introduced in this section therefore transfer less power from the lowest $k_\parallel$ (i.e. foreground) modes to the edge of the wedge.  Equivalently, if one defines the edge of the wedge by specifying a certain level of foreground contamination, our new estimators have caused the wedge to recede somewhat, giving a larger EoR window.

\begin{figure}[!ht] 
	\centering 
	\includegraphics[width=0.49\textwidth]{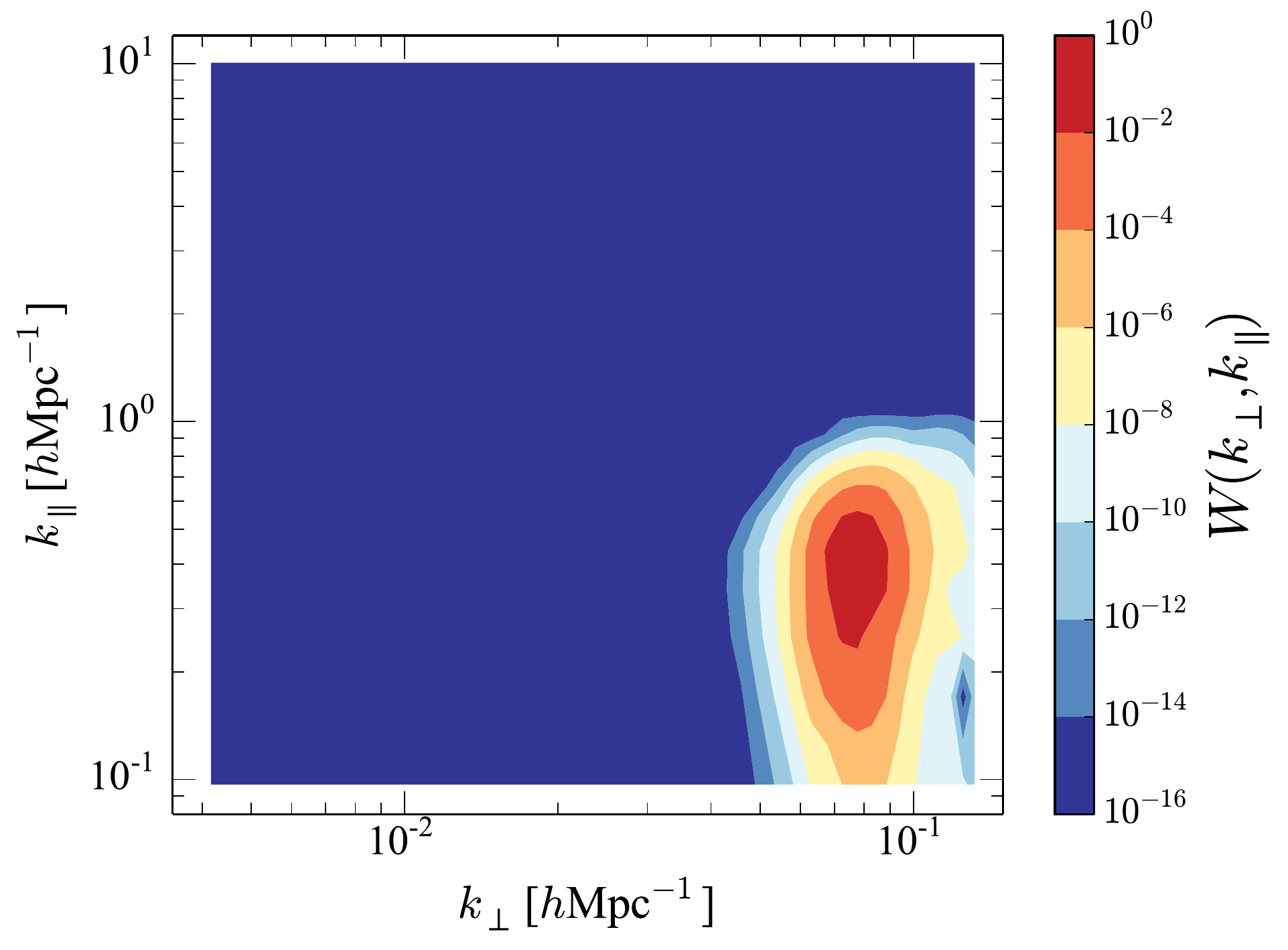}
	\caption{A window function from our optimal estimator, intended to probe $(k_\perp, k_\parallel) = (0.078, 0.17)\,h\textrm{Mpc}^{-1}$.  Rather than being centered on those coordinates, the window function instead peaks at $(k_\perp, k_\parallel) = (0.078, 0.44)\,h\textrm{Mpc}^{-1}$.  This results from the strong downweighting of foreground power at low $k_\parallel$ and the strong coupling between $k_\parallel$ modes at large $k_\perp$ (i.e. in the wedge region).}
	\label{fig:optEstOffset}
\end{figure} 

Examining the window function for a bandpower intended to probe a mode deep within the wedge reveals some unexpected behavior from our new estimators.  In Figure \ref{fig:optEstOffset} we show the window function for an optimal estimate of the power at $(k_\perp, k_\parallel) = (0.078, 0.17)\,h\textrm{Mpc}^{-1}$, which is well within the wedge seen in the previous section.  The window function is indeed sensitive to the mode that it was designed to probe, but its peak is at higher $k_\parallel$.  We find that the diagonal approximation gives essentially identical results.  Both the optimal estimator and its diagonal-inverse approximation heavily downweight foreground-contaminated modes at low $k_\parallel$, and when this is combined with an inherently strong instrumental coupling between different $k_\parallel$ modes, the result is a skewing of window functions towards higher $k_\parallel$.  To see this, consider a toy example where we attempt to make measurements at two different $k_\parallel$ modes---one low and one high---using just two delay bins from a single baseline.  The relationship between the input delays and the output $k_\parallel$ modes is encapsulated by the $\mathbf{C}_{,\alpha}$ matrices, and for our toy example, suppose we have
\begin{equation}
\mathbf{C}_{,\alpha} \Bigg{|}_{\alpha = 1} \equiv \frac{\partial \mathbf{C}\,\,}{\partial k^{\textrm{low}}_{\parallel}} = \left( \begin{array}{cc}
1 & 0.7 \\
0.7 & 0.6
\end{array}
\right)
\end{equation}
and
\begin{equation}
\mathbf{C}_{,\alpha} \Bigg{|}_{\alpha = 2} \equiv \frac{\partial \mathbf{C}\,\,}{\partial k^{\textrm{high}}_{\parallel}}  = \left( \begin{array}{cc}
0.6 & 0.7 \\
0.7 & 1
\end{array}
\right),
\end{equation}
where $k_\parallel^{\textrm{low}}$ and $k_\parallel^{\textrm{high}}$ are the two $k_\parallel$ values at which we wish to measure the power spectrum.  In the short baseline limit, Ref. \cite{Parsons2012b} showed that delay modes converge to $k_\parallel$ modes, and in such a limit the first matrix would contain just a single non-zero element in the top left corner, while the second matrix would similarly be non-zero only in its bottom left corner.  Here, we instead assume that we are dealing with a rather long baseline, since roughly speaking it is the long baselines of an array that probe high $k_\perp$, where the wedge resides.  For long baselines, delay modes are not a good approximation for $k_\parallel$ modes, and substantial non-zero values arise in other elements of the matrices, as our toy example illustrates.

Suppose now that $k_\parallel^{\textrm{low}}$ is at low enough $k_\parallel$ for the (true) bandpower $p_1$ to be foreground contaminated, while $k_\parallel^{\textrm{high}}$ is high enough for its bandpower $p_2$ to be essentially uncontaminated by foregrounds and instead dominated by the cosmological signal.  Taking $p_1 = 1$ and $p_2 = 0.01$ as an example, the covariance matrix $\mathbf{C}$ then becomes
\begin{equation}
\mathbf{C} =  \mathbf{N} + \sum_\alpha p_\alpha \mathbf{C}_{,\alpha} = \left( \begin{array}{cc}
1.061 & 0.77 \\
0.77 & 0.701
\end{array}
\right),
\end{equation}
where we have added an instrumental noise contribution $\mathbf{N} = \sigma^2 \mathbf{I}$, with $\sigma^2 = 10^{-3}$.  Inserting Eq. \eqref{eq:OptEalpha} for the optimal estimator into Eq. \eqref{eq:WindExplicitForm} with our matrices gives the window function matrix
\begin{equation}
\mathbf{W} =
\left( \begin{array}{cc}
0.37 & 0.63 \\
0.26 & 0.74
\end{array}
\right),
\end{equation}
where we have adhered to the convention established in Section \ref{sec:PaperI}, normalizing our estimator so that each row of the window matrix sums to unity.

Consider first the bottom row of $\mathbf{W}$.  This gives the linear combination of power that is probed by our estimator of the second, higher $k_\parallel$ mode.  We see that our estimator does indeed draw most of its power from the high $k_\parallel$ bin.  The top row of the matrix should ideally draw most of its power from the first $k_\parallel$ bin, but we see that this is not the case in our toy model.  We therefore see that the right combination of couplings in the $\mathbf{C}_{,\alpha}$ matrices, along with a lopsided spectrum of sky emission (with very strong emission in select foreground-contaminated modes) can cause the optimal estimator to skew the window functions away from their intended target modes.  Experimenting with different couplings and different foreground spectra in our toy model suggests that both conditions are necessary for this to happen.\footnote{Of course, our toy model is intended only for illustrative, proof-of-concept purposes.  However, it is encouraging to see that by setting up the same qualitative conditions as one has for the larger numerical computations (to the extent that one is able to using simple ``under-resolved" $2\times 2$ matrices), we are able to reproduce the rough behavior of the full results.}
In the full numerical computations, we find that the resulting skewing of window functions towards higher $k_\parallel$ happens for all the modes within the wedge.  Essentially, the optimal prescription is driving the sampling of Fourier modes away from the wedge to achieve its simple goal of minimizing the error bars.

\begin{figure*}[!ht] 
	\centering 
	\includegraphics[width=1.\textwidth]{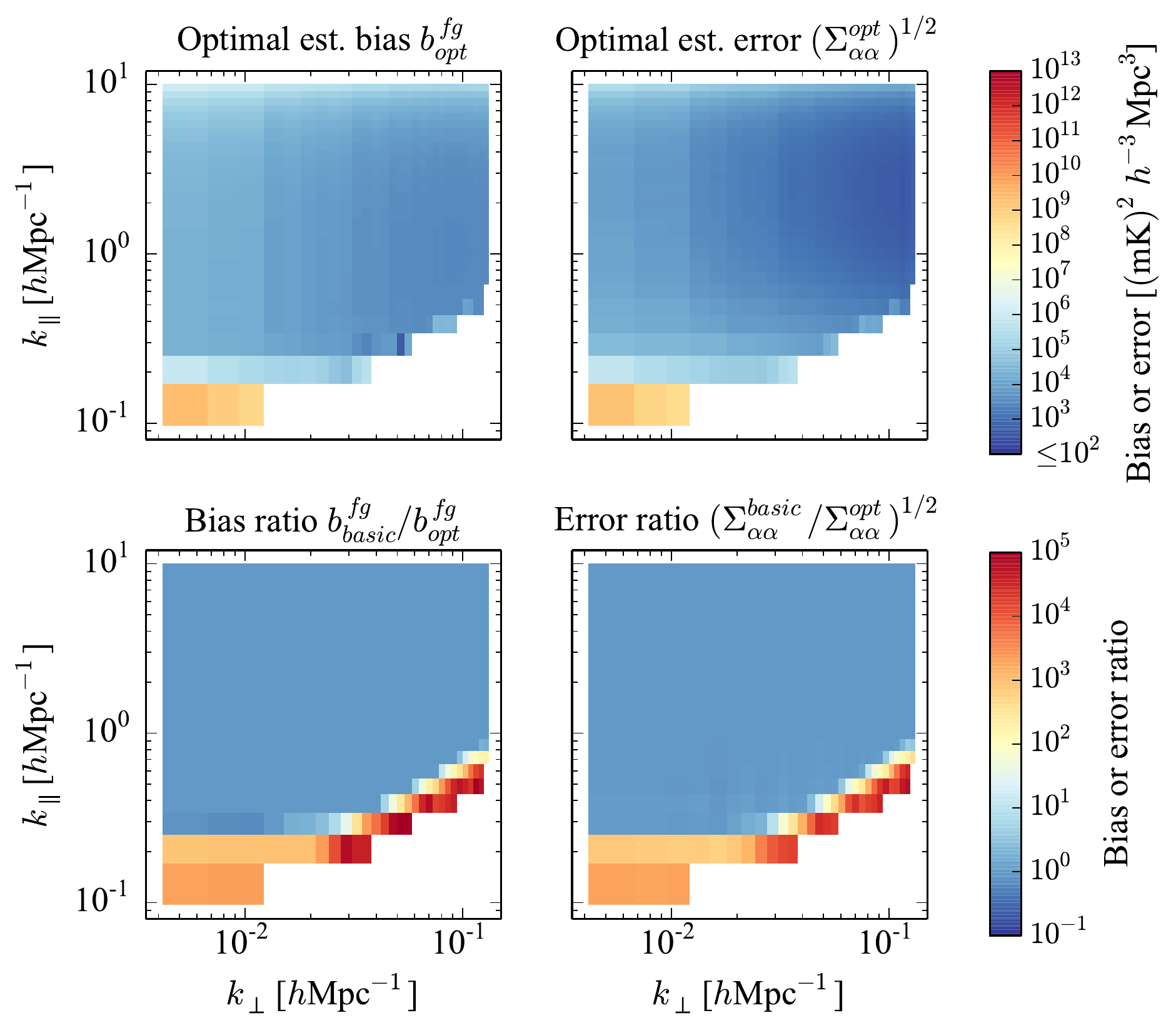}
	\caption{Rebinned error statistics for the optimal power spectrum estimator, taking into account the offsetting of window functions relative to their targeted bandpowers.  The left column shows the foreground bias using the optimal estimator, while the right column shows the power spectrum error bars.  This information is displayed in absolute terms along the top row, and as a ratio to the corresponding quantities with the basic estimator along the bottom row.  The use of an optimal estimator allows one to expand the EoR window to lower $k_\parallel$, where low $k$ measurements of the power spectrum can be made.}
	\label{fig:optBiasAndErrorComparisons}
\end{figure*} 

\begin{figure}[t] 
	\centering 
	\includegraphics[width=0.49\textwidth]{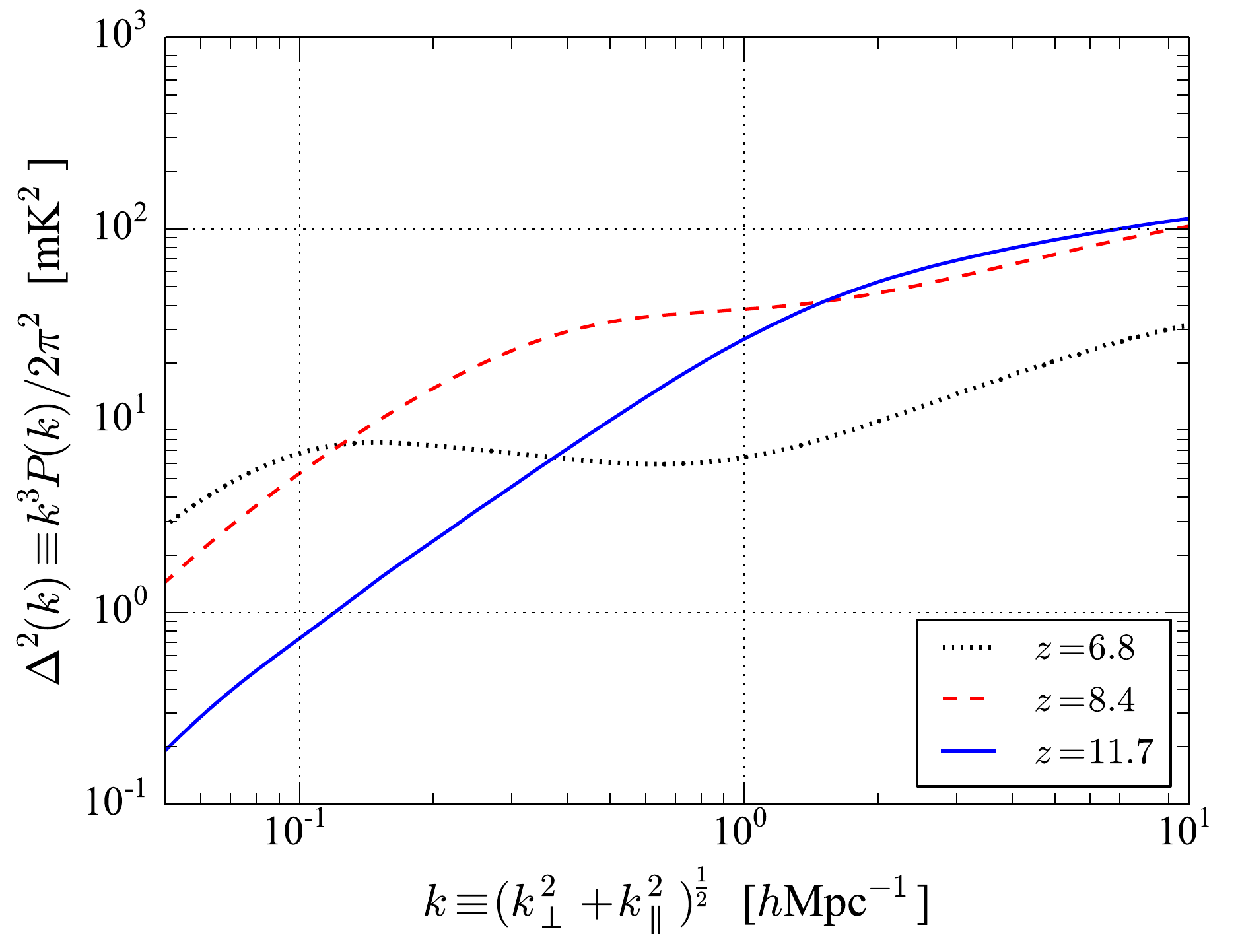}
	\caption{Example power spectra from Ref. \cite{Barkana2009}, plotted as $\Delta^2(k) \equiv \frac{k^3}{2\pi^2} P(k)$.  The use of the optimal power spectrum estimator allows the detection of the cosmological signal to rise from $2.4\sigma$ to $5\sigma$ at $z\sim 11.7$, from $12\sigma$ to $50\sigma$ at $z\sim 8.4$, and from $5.6\sigma$ to $35\sigma$ at $z \sim6.8$.}
	\label{fig:sampleTheory}
\end{figure}

Having noted that our power estimates may not necessarily peak at their intended location on the $k_\perp k_\parallel$ plane, we may use the window functions to dictate a re-binning of our power estimates (and their corresponding error statistics).  We do so using the binning formalism of Ref. \cite{Dillon2014}, placing all the relevant quantities at their correct locations in Fourier space.  The resulting error bars and foreground biases are shown in Figure \ref{fig:optBiasAndErrorComparisons}, along with the ratios of these quantities to their counterparts with the basic estimator of Paper I and Section \ref{sec:PaperI}.  In terms of both the foreground bias and the error bars, the use of the optimal estimator allows one to access the lowest $k$ modes of the power spectrum.  At the boundaries of the EoR window, one can typically reduce the biases and errors by a factor of a few to $10^5$, with the greatest gains typically occurring at high $k_\perp$ near the wedge.  Given that the cosmological signal-to-instrumental noise ratio increases quickly towards lower $k$, this results in a large increase in detection significance.  Suppose, for instance, we take the simulations of Ref. \cite{Barkana2009} as a fiducial model of reionization.  These simulations reach an ionization fraction of slightly under $50\%$ at the redshift of our numerical computations ($z \sim 8.4$).  Extracting the relevant power spectrum (shown in Figure \ref{fig:sampleTheory}) and assuming isotropy to compare it to the $k_\perp k_\parallel$ space error computations shown in Figure \ref{fig:optBiasAndErrorComparisons}, we find that increasing the size of the EoR window with the optimal estimator results in an increase in the detection significance of the cosmological signal from $12\sigma$ to $50\sigma$.  To get a rough sense for this enhancement in performance at different redshifts, we may rescale our computed errors using $T_\textrm{sky} \propto \nu^{-2.55}$ \cite{thompson_et_al_2007}, since both the foreground uncertainty and the instrumental noise (which is sky noise dominated) scale with the sky temperature $T_\textrm{sky}$.  This treatment is of course only approximate, but saves us from redoing the rather costly numerical computations at multiple redshifts, and suffices for a back-of-the-envelope estimate.  At $z \sim 6.8$ (near the end of reionization for this model), the detection significance is boosted from $5.6\sigma$ to $35\sigma$.  At $z \sim 11.7$ (near the beginning of reionization in this model), higher noise and brighter foregrounds make detections more difficult, but the optimal estimator allows a marginal $2.4\sigma$ detection to become a statistically significant $5\sigma$ detection.  While these precise numbers are of course dependent on observational details (such as the configuration of one's array and the integration time, which in our case are taken from Paper I), they do suggest that the optimal estimator's slight enlargement of the EoR window can significantly boost detection significance.

Away from the most foreground-contaminated regions, we can see from Figure \ref{fig:optBiasAndErrorComparisons} that the basic estimator essentially performs just as well as the optimal estimator does.  This is unsurprising, given that the window functions look rather similar to each other away from the wedge in Figure \ref{fig:comparisonWindCollection}.  Another way to understand this is to recall that the basic estimator is extremely similar to the FKP estimator in the small scale (high Fourier wavenumber) limit far away from the wedge, and the FKP approximation is known to be optimal in that limit.  We find that throughout most of the EoR window (where thermal noise dominates the errors), the final error bars on power spectrum measurements typically differ by less than a few percent between the estimators.

When comparing the optimal estimator to its diagonal-inverse approximation, we find that the agreement is even better, in that one need not be safely inside the EoR window for the diagonal-inverse approximation to work well.  The percent-level agreement extends right down to the peripheries of the EoR window, which perhaps explains the excellent recent progress in delay-based foreground isolation techniques \cite{Parsons2013}.

That the basic estimator and the diagonal-inverse approximation work so well is good news from the perspective of computational resources.  First, it suggests that for those who wish to be maximally cautious about foreground systematics by working only in the EoR window, the basic estimator (which requires no foreground modeling whatsoever) is optimal for all intents and purposes.  Moreover, those who wish to push to the edges of the EoR window can do so by replacing the $\mathbf{C}^{-1}$ part of the optimal estimator with its diagonal approximation in the delay basis.  The normally costly matrix inversion then becomes trivial, and since the delay transform acts on a per-baseline basis, the transform itself does not constitute a burdensome computation.

\section{Accessing the wedge via window function decorrelation}
\label{sec:Decorr}
In the previous section, we saw that using an optimal estimator allowed one to slightly expand the size of the EoR window.  We will now introduce extensions to this approach that may---if numerical instabilities can be well controlled---allow further enlargements of the EoR window into the foreground wedge.

To guide our next steps, consider the weaknesses of our optimal estimator thus far:
\begin{enumerate}
\item[(1)] Because the data covariance matrix $\mathbf{C}$ enters as part of the power spectrum estimator itself (in contrast to our basic estimator, where $\mathbf{C}$ is needed only to determine the final error properties), our estimator is suboptimal if $\mathbf{C}$ is incorrectly modeled.  A mis-modeling of $\mathbf{C}$ (for example, due to systematics in our knowledge of foregrounds) will impact our ability to suppress foregrounds.  While formally speaking, perturbations to $\mathbf{C}$ give rise only second-order effects \cite{Tegmark2001}, the large amplitude of foregrounds makes such effects potentially important.
\item[(2)] Even if the covariance is perfectly modeled, our estimator will be optimal only in the limit of Gaussian fluctuations.  Foregrounds are known to be non-Gaussian in their distribution, although the impact of this non-Gaussianity on power spectrum estimation has yet to be fully quantified.
\item[(3)] The optimal estimator essentially gives up on estimating Fourier modes that are deep within the foreground wedge, potentially lowering the significance of a power spectrum measurement.
\end{enumerate}

We begin by tackling the last problem, which will eventually lead us to solutions for the first two problem as a bonus.  From the last few sections, we have seen that the phenomenon of the foreground wedge can be thought of in terms of window functions, which act as position-dependent point spread functions on the Fourier plane and scatter foregrounds from their ``native" low $k_\parallel$ region into the wedge.  With our covariant description, the window functions are precisely quantified, which means that their effect can in principle be undone.  From Eq. \eqref{eq:WindDef}, we see that this can be accomplished by taking our estimated bandpowers and multiplying by the inverse window function matrix $\mathbf{W}^{-1}$:
\begin{equation}
\hat{p}_\alpha^\prime = \sum_{\beta} \mathbf{W}^{-1}_{\alpha \beta} \widehat{p}_\beta,
\end{equation}
where $\hat{p}_\alpha^\prime$ is the $\alpha$th bandpower of our new deconvolved estimator.\footnote{We emphasize that the deconvolution described in this section is a post-processing step that acts in the ``output space" of bandpowers after an initial power spectrum has been estimated.  It is not to be confused with the more conventional use of the term ``deconvolution" in radio astronomy, which typically refers to the fitting out of bright point sources in images prior to power spectrum estimation.}  The error covariance is correspondingly revised to be
\begin{equation}
\boldsymbol \Sigma^\prime =  \mathbf{W}^{-1} \boldsymbol \Sigma (\mathbf{W}^{-1})^t.
\end{equation}
By construction, the window function matrix is now given by the identity matrix, since $\mathbf{W}^{-1} \mathbf{W} = \mathbf{I}$.  This is equivalent to saying that our new estimator has no leakage of power between different cells on the $k_\perp k_\parallel$ plane.  Importantly, this means that foregrounds remain in the lowest $k_\parallel$ modes.  From a mathematical standpoint, then, the foreground wedge need not exist at all, and one should in principle be able to work anywhere where the foregrounds are not intrinsically bright.

In some ways, we have accomplished very little with our new scheme.  While it may be cosmetically attractive to banish the leaked foregrounds from the wedge, multiplying our bandpowers by $\mathbf{W}^{-1}$ does not change the information content of our measurement, since $\mathbf{W}^{-1}$ is an invertible matrix.  If we stopped here, any science results from our power spectrum measurement (such as any parameter constraints on theoretical models) would yield identical results whether or not we chose to deconvolve the window functions.

The deconvolution of the window functions, however, opens the door to one more round of foreground subtraction.  In Ref. \cite{Shaw2014b} it was emphasized that the chromatic nature of an interferometer---the cause of the wedge---does not necessarily make the foregrounds inherently worse.  They are simply more difficult to identify because they have leaked into Fourier modes where they would not be naively expected.  Any additional foreground removal at this stage would therefore require the quantification of this leakage for a large number of modes (essentially all the modes inside the foreground wedge).  With a deconvolved estimate of the power spectrum, the foregrounds are confined to a smaller set of modes, which can then be projected out of our measurement.  One could, for example, multiply the deconvolved bandpowers by a projection matrix that zeros out all bandpowers at the lowest $k_\parallel$ values of our estimate, since those correspond to sky contributions that are the most spectrally smooth, and therefore most likely to be dominated by foregrounds.  Such a matrix would be given by
\begin{equation}
\boldsymbol \Pi = \left( \begin{array}{ccc|ccccc}
0&   & && & &&\\
& 0 && & &&&\\
    &  & \ddots & & &&&\\
  \hline
&  & &1&  &  & &  \\
&  & &  &1 & & &  \\
&  & &  &  &1 & &  \\
&  & & & &&1&\\
    &  & & & &&&\ddots\\
\end{array}
\right),
\end{equation}
where blank portions of the matrix are zero, and we have assumed that the matrix elements are ordered in such a way that one cycles through the different $k_\perp$ values more rapidly than the $k_\parallel$ values, so that a projection matrix eliminating the $N$ lowest $k_\parallel$ values is formed by zeroing out the first $N \times N_{k_\perp}$ rows and columns of an identity matrix, where $N_{k_\perp}$ is the number of different values of $k_\perp$ on our Fourier grid.

By projecting out the lowest (and presumably most foreground contaminated) $k_\parallel$ mode(s), we have immunized ourselves against any foreground modeling uncertainties in those modes, solving the first problem enumerated above.  The approach suggested here therefore stands in contrast to the optimal methods used in the previous section, which depended on having a full foreground model.  In spirit, our new scheme is quite similar to foreground avoidance strategies that simply accept the existence of the wedge and look elsewhere in Fourier space.  Here, we are simply not looking at parts of the Fourier plane that are intrinsically foreground contaminated following window function deconvolution.  But with foreground leakage into the wedge undone by the deconvolution, our scheme requires fewer Fourier modes to be ignored.  This is desirable from the standpoint of sensitivity.  To see this, recall that foregrounds are typically much brighter in amplitude than the cosmological signal.  This means that even small fractional levels of foreground leakage into a Fourier cell effectively render the cell unusable.  (This was the motivation for delay-spectrum isolation techniques that use the horizon to establish a relatively sharp cut-off on smooth spectrum foreground leakage \cite{Parsons2012b,Parsons2013}).  A large number of Fourier cells are therefore wasted in a traditional wedge-avoidance strategy, where it is simply accepted that window functions in the wedge are broad and that leakage is commonplace.  Window function deconvolution limits the number of Fourier modes that must be discarded, sequestering foregrounds to Fourier cells where the foregrounds intrinsically reside (and are therefore irretrievably lost anyway), allowing greater signal-to-noise in other regions.

Importantly, we emphasize that it is possible to isolate and project out the low $k_\parallel$ modes whether or not the underlying fluctuations are Gaussian.  This is because Eq. \eqref{eq:WindExplicitForm} for the window function matrix depends only on two-point statistics.  The window function matrix therefore accurately describes the relationship between the true power spectrum and our measured one.  This allows the low $k_\parallel$ modes to be isolated even if those fluctuations are non-Gaussian.  Once those modes have been projected out, it is irrelevant whether the foregrounds there were non-Gaussian or not, since they no longer have any impact on our measurement, thus solving the second problem on our list.

As discussed previously in the literature (in Refs. \cite{Dillon2014,Shaw2014b}, for example) and summarized briefly in Appendix \ref{appendix:Fisher}, one disadvantage of fully deconvolving out the window functions is that neighboring bandpower estimates end up having negatively correlated errors.  This arises because we have essentially demanded more information than our survey has had to offer: after deconvolution, the window functions are formally delta functions, which is not ``naturally" achievable unless an infinite volume is surveyed.  One approach to this problem is to resmooth the data following our deconvolution and projection.  In what follows we will demonstrate how this can be achieved while maintaining our method's insensitivity to the modes that have been projected out.

Consider the power spectrum error covariance matrix immediately following the window function deconvolution and projection:
\begin{equation}
\boldsymbol \Sigma^{\prime \prime} = \boldsymbol \Pi  \mathbf{W}^{-1} \boldsymbol \Sigma \mathbf{W}^{-t} \boldsymbol \Pi.
\end{equation}
If one were to apply a resmoothing/convolution matrix $\mathbf{G}$ to the data, the new error covariance would be $\mathbf{G} \boldsymbol \Sigma^{\prime \prime} \mathbf{G}^t$.  The matrix $\mathbf{G}$ can be chosen arbitrarily, but a particularly attractive choice would be one that results in a diagonal error covariance, so that the measured bandpowers have uncorrelated errors.\footnote{In the language of Appendix \ref{appendix:Fisher}, we would like to emulate the spirit of the $\mathbf{M} \propto \mathbf{F}^{-1/2}$ normalization for the power spectrum estimator while preserving the gains that we have made in foreground mitigation.}  If $\boldsymbol \Sigma^{\prime \prime}$ were a full rank matrix, one could simply pick $\mathbf{G} \propto (\boldsymbol \Sigma^{\prime \prime})^{-\frac{1}{2}}$.  Such a choice is impossible, however, because our elimination of the lowest $k_\parallel$ modes makes $\boldsymbol \Sigma^{\prime \prime}$ singular, with the first few rows and columns identically zero.  Fortunately, the submatrix $\boldsymbol \Sigma^{\prime \prime}_\textrm{sub}$ away from those zeroed rows and columns is still non-singular, and it is possible to instead use
\begin{equation}
\mathbf{G} \propto \left( \begin{array}{ccc|ccc}
0&   & && & \\
& 0 && & &\\
    &  & \ddots & & &\\
  \hline
&  & &&  &   \\
&  & &  & (\boldsymbol \Sigma^{\prime \prime}_\textrm{sub})^{-\frac{1}{2}} &   \\
&  & &  &  &  \\
\end{array}
\right).
\end{equation}
Putting this all together, then, our latest estimator (termed the ``foreground isolation estimator" $\mathbf{\widehat{p}}^\textrm{iso}$) takes the form
\begin{equation}
\mathbf{\widehat{p}}^\textrm{iso} \equiv \mathbf{G} \boldsymbol \Pi \mathbf{W}^{-1} \mathbf{\widehat{p}},
\end{equation}
where $\mathbf{\widehat{p}}$ is a power spectrum estimate from the previous section.  (We find in our computations that one obtains essentially the same results whether $\mathbf{\widehat{p}}$ is estimated using the optimal estimator or its diagonal-inverse approximation).  The normalization of $\mathbf{G}$ is once again determined by requiring the rows of the window function matrix sum to unity, which is now given by
\begin{equation}
\label{eq:IsoWindows}
\mathbf{W}^\textrm{iso} = \mathbf{G} \boldsymbol \Pi.
\end{equation}

Before we examine some numerical results from an application of our foreground isolation estimator, let us briefly compare and contrast the foreground projection proposed here and direct foreground subtraction algorithms that have been previously considered in the literature.  Earlier work often focused on working directly with input data such as sky maps, subtracting off smooth spectral components that are believed to be foreground contaminated.  These smooth components are often modeled as low-order polynomials \cite{Wang2006,Liu2009a,Bowman2009,Liu2009b} or as long-wavelength line-of-sight Fourier modes \cite{Petrovic2011}.  The latter choice is particularly similar to our foreground isolation approach.  However, direct subtraction methods raise the concern of cosmological signal loss, making it crucial to simulate and propagate the effects of subtraction on the signal.  This is particularly important because such effects are rarely perfectly localized in Fourier space once the data have been pushed through the entire power spectrum estimation pipeline.  In contrast, the foreground isolation approach is guaranteed, by construction, to affect only the chosen $k_\parallel$ modes.  In addition, note from Eq. \eqref{eq:IsoWindows} that all window functions (i.e. all rows of $\mathbf{W}^\textrm{iso}$) are identically zero in regions that have been projected out.  There is thus never any formal loss of cosmological signal, since our window functions reflect the reality that we have not made any measurements in regions that have been discarded.

\begin{figure*}[!ht] 
	\centering 
	\includegraphics[width=1.\textwidth]{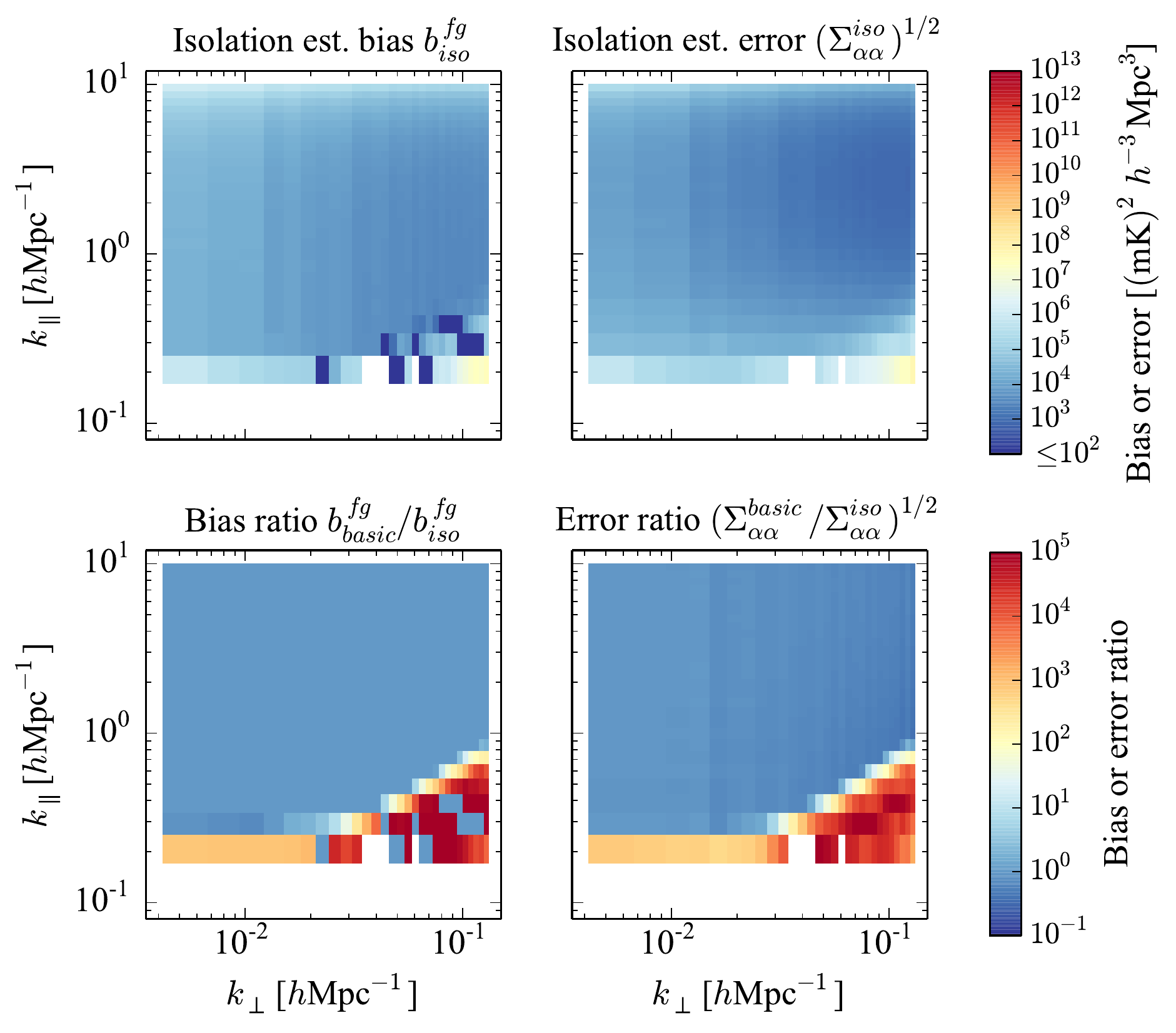}
	\caption{Similar to Figure \ref{fig:optBiasAndErrorComparisons}, but for the foreground isolation estimator instead of the optimal estimator.  One sees that the foreground isolation estimator potentially allows one to work within the wedge, provided numerical artifacts can be controlled.}
	\label{fig:isolationBiasAndErrorComparisons}
\end{figure*} 

In Figure \ref{fig:isolationBiasAndErrorComparisons}, we compare the biases and errors of the foreground isolation estimator to those of the basic estimator.  In contrast to the estimators of Section \ref{sec:BetterEst}, the foreground isolation approach allows one to operate within the wedge, with errors and biases there reduced by up to five orders of magnitude compared to those of the basic estimator.  That a small ``shadow" of the wedge can still be seen is due to two reasons.  First, our foreground isolation estimator has decorrelated error bars, so that the errors on each individual $k_\perp k_\parallel$ cell are independent.  In Paper I, we saw that a chromatic instrument tends to produce correlated measurements in the wedge, and undoing such correlations will necessarily increase the size of the errors.  This also happens outside the wedge, with the decorrelated errors being on average $1.4$ times larger (and at worst $2.6$ times larger) than the errors from both the basic and the optimal estimators (which, recall, perform roughly equally well outside the wedge).  Such error increases are not a cause for concern as they simply reflect the underlying reality of error correlations.  If desired, the errors can be brought back down by re-binning, with no change in the statistical significance of the final power spectrum measurement

Of more serious concern is the possibility of numerical errors.  Indeed, there is evidence of such issues in our computations.  In the bias plots of Figure \ref{fig:isolationBiasAndErrorComparisons}, for example, small patches of anomalous cells are visible in the middle of what is otherwise a smooth background.  The patches represent bandpowers that are estimated to be negative following the foreground isolation process.  These are caused by the algorithm's attempts to subtract two large foreground contributions within the wedge to yield a small residual.  While this is in principle possible, it is of course prone to numerical errors, which in our case is concretely attributable to small errors in the application of $\mathbf{W}^{-1}$.  These errors are apparent if one examines the window functions immediately following this step.  In principle, the window functions should be delta functions.  We find this to be mostly the case, but windows centered inside the wedge possess numerical noise at the $10^{-6}$ level away from the peak.  While such a level of error may be sufficient for many applications, it is worrying for our present one, given the large dynamic range between the amplitudes of the foregrounds and the instrumental noise.

That numerical issues exist with this approach is not an artifact of our implementation, and is in fact expected.  As we have seen in previous sections, the window functions within the wedge are typically either off-centered and encroach on regions normally associated with other bandpowers (as is the case with the optimal estimator) or are severely elongated (as is the case with the basic estimator).  Many of the windows are therefore very similar in shape and location, resulting in nearly identical rows---and therefore near-singularity---in the window function matrix.  Inverting this matrix therefore requires teasing out tiny differences between the wings of different window functions, which is clearly an operation that requires very high numerical precision.  Despite this, the attractive properties of the foreground isolation approach (namely the formal preservation of the cosmological signal and the foreground-model independence in the discarded modes) make the numerical challenges worth overcoming in future work.

\section{Conclusions}
\label{sec:Conclusions}
This paper is a direct continuation of Paper I \cite{Liu2014a}, where a rigorous framework was established for describing the foreground wedge and its complement, the EoR window.  In this paper, we made use of our framework to examine statistical methods for enlarging the EoR window, with an eye towards enabling high significance measurements of the power spectrum deep inside the foreground wedge.

In Paper I we examined a basic noise-weighted estimator of the power spectrum.  In the limit of a very finely discretized Fourier plane, we showed that this is equivalent to a separable estimator where the visibilities are gridded with a primary beam kernel, and then squared and binned.  In Section \ref{sec:FKP} of the present paper we considered a more general class of separable estimators, and searched for a more optimal way to weight our data by adapting the FKP approximation from galaxy surveys.  Assuming that the power spectrum is a slowly varying function and that one is working on small scales, the FKP approximation can be shown to be optimal (even when compared to the more general class of non-separable estimators).  It is therefore a useful reference for comparing with our basic prescription.  Making such a comparison reveals that the basic prescription is optimal only if the thermal noise is large compared to the sky signal.  Such an assumption is clearly violated as one moves from higher $k_\parallel$ to lower $k_\parallel$ in an attempt to work within the wedge.  This can be rectified by weighting the data using FKP weights instead of just instrumental noise.  However, as one pushes to the lowest $k_\parallel$, the assumptions underlying the FKP approximation itself (such as the assumption that one is working on small length scales) break down.  While the FKP approximation may be a useful approximation that is relatively computationally cheap, a viable prescription for working within the wedge will involve more complicated non-separable estimators.

In Section \ref{sec:BetterEst}, we considered the larger class of non-separable estimators.  Performing a numerical analysis of what is a provably optimal estimator, we find that the optimal estimator's window functions are naturally shifted away from the deepest depths of the wedge.  However, the optimal estimator is able to reduce the foreground biases and errors by up to factors of $10^5$ in power at the edge of the wedge, thus enlarging the EoR window slightly.  In our fiducial models, this boosts the detection significance of the cosmological signal from $12\sigma$ to $50\sigma$ around the midpoint of reionization, although this is of course an array-dependent statement.

With a fully covariant framework (and thus the ability to compute window functions), one can trade ``horizontal error bars" for ``vertical error bars" in a power spectrum measurement.  By reducing the horizontal error bars (i.e., by narrowing the window functions), one can reduce the leakage of foregrounds across the Fourier plane, which was what caused the foreground wedge in the first place.  In Section \ref{sec:Decorr}, we proposed a foreground isolation scheme where the window functions are first fully deconvolved to delta functions in order to isolate the foregrounds as much as possible.  This is followed by a zeroing out of modes that still contain strong foreground contamination, before a final resmoothing (or rebinning) of the $k_\perp k_\parallel$ plane to reduce the vertical error bars, which were inadvertently magnified by the window function deconvolution.  While future work is required to tame expected numerical issues, early indications suggest that this scheme may allow measurements to be made within the wedge.  Crucially, our proposed method has no formal signal loss associated with it, and does not assume that the sky signals are Gaussian.

Importantly, we note that the techniques developed in this paper are widely applicable to cosmological $21\,\textrm{cm}$ surveys at all redshifts, not just those that seek to study reionization.  Our conclusions rely only on the assumptions that one uses an interferometer to map the intensity of a spectral line, and that the foregrounds are spectrally smooth to some degree.  Indeed, techniques similar to the ones we explore in Paper I and the present paper have been applied to forecasts for lower-redshift $21\,\textrm{cm}$ surveys targeting baryon acoustic oscillation measurements \cite{Shaw2014a,Shaw2014b}.

In battling the twin challenges of sensitivity and foreground systematics, $21\,\textrm{cm}$ power spectrum measurements are in a somewhat unfortunate situation, with thermal noise large where foregrounds are weak and vice versa.  To increase the significance of future measurements, one must therefore either reduce the thermal noise or the influence of foregrounds in the final power spectrum.  These approaches are of course not mutually exclusive, and next-generation instruments such as the recently-proposed HERA \cite{Pober2014} and the SKA will achieve the former, while in this paper we have proposed various methods for achieving the latter.  The work presented here suggests that it may indeed be possible to enlarge the EoR window, and further progress on this front will be a crucial step in fully realizing the great promise of $21\,\textrm{cm}$ cosmology.

\section*{Acknowledgments}
We would like to thank Gianni Bernardi, Chris Carilli, Josh Dillon, Michael Eastwood, Aaron Ewall-Wice, Danny Jacobs, Matt McQuinn, Miguel Morales, Abraham Neben, Peter Nugent, Jonnie Pober, Jonathan Pritchard, Richard Shaw, Casey Stark, Max Tegmark, and Chris Williams for useful discussions.  We would also like to acknowledge the anonymous referee of this paper for extremely constructive comments.  Additionally, we thank the anonymous referee of Ref. \cite{Dillon2014} for encouraging the resumption of a previously-abandoned line of thought that eventually led to Section \ref{sec:Decorr}.  
This research used resources of the National Energy Research
Scientific Computing Center, which is supported by the Office of
Science of the U.S. Department of Energy under Contract No. 
DE-AC02-05CH11231.  The Centre for All-sky Astrophysics is an Australian Research Council Centre of Excellence, funded by grant CE110001020. The International Centre for Radio Astronomy Research is a Joint Venture between Curtin University and the University of Western Australia, funded by the State Government of Western Australia and the Joint Venture partners.  This work was partially supported by the National Science Foundation (awards 0804508, 1129258, and 1125558), and a generous grant from the Mt. Cuba Astronomical Association.

\appendix
\section{Fisher Matrix Formalism}
\label{appendix:Fisher}

In this Appendix, we briefly discuss the Fisher matrix formalism and its applications to the estimation of power spectra.  Let $\boldsymbol \theta$ be a set of parameters that we wish to constrain, and let $L(\mathbf{x}; \boldsymbol \theta)$ be its likelihood function given data $\mathbf{x}$ from our experiment.  The Fisher matrix $\mathbf{F}$ measures the information content of our experiment, and its elements are given by
\begin{equation}
\mathbf{F}_{\alpha \beta} = - \Bigg{\langle} \frac{\partial^2 \ln L}{\partial  \theta_\alpha  \partial  \theta_\beta} \Bigg{\rangle}.
\end{equation}
Typically, one wishes to minimize the variance on the estimator $\hat{\boldsymbol \theta}$ of our parameters, which is given by the diagonal of the covariance matrix $\boldsymbol \Sigma$, where
\begin{equation}
\boldsymbol \Sigma \equiv \langle \hat{\boldsymbol \theta} \hat{\boldsymbol \theta}^\dagger \rangle - \langle \hat{\boldsymbol \theta} \rangle \langle \hat{\boldsymbol \theta} \rangle^\dagger.
\end{equation}
If we have an estimator with no multiplicative bias, such that $\langle \hat{\boldsymbol \theta} \rangle= \boldsymbol \theta$, the Cramer-Rao inequality states that the resulting error bar on $\Delta \theta_\alpha$ on the parameter $\theta_\alpha$ must satisfy $\Delta \theta_\alpha \ge (\mathbf{F}^{-1})_{\alpha \alpha}$.  The Fisher matrix therefore provides a hard limit on the smallest possible error bars for an unbiased estimator, and an optimal unbiased estimator is one that saturates the Cramer-Rao inequality.

In this paper our focus is on estimating power spectrum bandpowers.  For us, the parameter vector $\boldsymbol \theta$ is therefore the bandpower vector $\mathbf{p}$, and assuming a Gaussian form for the likelihood yields
\begin{equation}
\mathbf{F}_{\alpha \beta} = \frac{1}{2} \textrm{tr} \left[ \mathbf{C}^{-1} \mathbf{C}_{,\alpha} \mathbf{C}^{-1} \mathbf{C}_{,\beta}\right],
\end{equation}
where (like in the main text) $\mathbf{C} \equiv \langle \mathbf{x} \mathbf{x}^\dagger \rangle$, $\mathbf{C}_{,\alpha} \equiv \partial \mathbf{C}/ \partial p_\alpha$, and we have assumed that $\langle \mathbf{x} \rangle = 0$.  Suppose that we take the following form as an ansatz for an optimal, unbiased power spectrum estimator:
\begin{equation}
\label{eq:EstAnsatz}
\widehat{p}_\alpha = \sum_{\beta} M_{\alpha \beta} \mathbf{x}^\dagger \mathbf{C}^{-1} \mathbf{C}_{,\alpha} \mathbf{C}^{-1} \mathbf{x}.
\end{equation}
Taking the expectation value of this expression and inserting Eq. \eqref{eq:CovarDecomp} gives
\begin{equation}
\langle \widehat{\mathbf{p}} \rangle = \mathbf{M} \mathbf{F} \mathbf{p} + \mathbf{b},
\end{equation}
where $\mathbf{b}$ is the additive bias term, given by Eq. \eqref{eq:bias}.  The additive bias can be modeled and subtracted from the estimator, or (as we do in this paper) be left untouched as a systematic that is hopefully small in the parts of Fourier space of interest, such as the EoR window.  Comparing the rest of the equation with Eq. \eqref{eq:WindDef}, we see that $\mathbf{W} = \mathbf{M}\mathbf{F}$.  Computing the covariance of our bandpower estimate yields
\begin{equation}
\boldsymbol \Sigma = \mathbf{M}\mathbf{F}\mathbf{M}^t.
\end{equation}

Suppose we pick $\mathbf{M} = \mathbf{F}^{-1}$.  We then have $\mathbf{W} = \mathbf{I}$, so that the estimator has no multiplicative bias and satisfies $\langle \widehat{\mathbf{p}} \rangle = \mathbf{p}$ (where we have omitted the additive bias term for notational brevity as it is irrelevant to the discussion that follows).  We also have $\boldsymbol \Sigma = \mathbf{F}^{-1}$, so the error bars on our bandpowers saturate the Cramer-Rao bound.  The estimator given by Eq. \eqref{eq:EstAnsatz} with the choice $\mathbf{M} = \mathbf{F}^{-1}$ is therefore the optimal, unbiased estimator.

Using the optimal unbiased estimator in practice, however, one tends to get rather large error bars.  To understand this, recall that the optimal unbiased estimator is a solution to a \emph{constrained} minimization problem: it is the estimator that delivers the smallest error bars within the requirement that $\mathbf{W} = \mathbf{I}$.  This requirement is rather restrictive, as it essentially mandates ``horizontal error bars" of zero width.  The result is a larger set of ``vertical error bars".  In addition, neighboring off-diagonal elements of the error covariance matrix $\boldsymbol \Sigma$ will tend to have opposite signs, resulting in anti-correlated errors.

An alternative to the $\mathbf{M} = \mathbf{F}^{-1}$ estimator is one where $\mathbf{M}$ is taken to be diagonal, with each diagonal element fixed by the requirement that the window functions be correctly normalized (e.g., by requiring that rows of the $\mathbf{W}$ sum to unity).  This choice yields a non-identity window function matrix.  In other words, the power spectrum estimator, though correctly normalized, possess a multiplicative bias (in the matrix sense).  Free from the restriction of being unbiased, the diagonal $\mathbf{M}$ estimator is able to deliver error bars that are smaller than those dictated by the Cramer-Rao bound, and in fact are the smallest possible of any estimator \cite{Tegmark1997b}.  This estimator is what we refer to in the main text as the ``optimal, minimum-variance estimator" or ``optimal estimator" for short.  However, it should be noted that this estimator tends to give errors that are correlated between neighboring bandpowers.

A third option is to pick $\mathbf{M} \propto \mathbf{F}^{-\frac{1}{2}}$, with the normalization of each row determined again by the requirement that the window functions be normalized.  The error bars are slightly larger than those of the minimum-variance estimator, but have the virtue of being uncorrelated, since $\boldsymbol \Sigma$ is now diagonal.

The various approaches to EoR window enlargement that are explored in this paper can be traced back to the three options for $\mathbf{M}$ that we have outlined here.  The optimal estimator discussed in Section \ref{sec:BetterEst} is the minimum-variance estimator where $\mathbf{M}$ is taken to be diagonal.  The foreground isolation strategy discussed in Section \ref{sec:Decorr} is a combination of the unbiased $\mathbf{M} = \mathbf{F}^{-1}$ and the uncorrelated option with $\mathbf{M} \propto \mathbf{F}^{-\frac{1}{2}}$: the unbiased estimator is used first to give the sharpest window functions ($\mathbf{W} = \mathbf{I}$), allowing the foregrounds to be isolated to the lowest $k_\parallel$ regions, where they are then projected out before the bandpowers are re-smoothed to give the equivalent of the uncorrelated estimator within the remaining subspace.

\bibliography{wedgeFormalismPartB}

\end{document}